\documentclass[conference]{IEEEtran}

\usepackage{comment}
\usepackage{paralist}
\usepackage{subcaption}
\usepackage{graphicx}
\usepackage{amsmath}
\usepackage{url}
\usepackage{color}

\newcommand{\sender}{{\ensuremath{\sf Snd}}}
\newcommand{\receiver}{{\ensuremath{\sf Rcv}}}
\newcommand{\router}{{\ensuremath{\sf Rt}}}
\newcommand{\producer}{{\ensuremath{\sf Pr}}}
\newcommand{\adversary}{{\ensuremath{\sf Adv}}}
\newcommand{\snr}{{\sender\ and \receiver}}
\newcommand{\expiration}{{\ensuremath Exp_\router}}

\newcommand{\hit}{{\ensuremath \mbox{RTT}_{\sf hit}}}
\newcommand{\miss}{{\ensuremath \mbox{RTT}_{\sf miss}}}

\newcommand{\interest}[2]{{\ensuremath [I : #1 \rightarrow #2]}}
\newcommand{\content}[2]{{\ensuremath [C : #1 \rightarrow #2]}}

\usepackage{balance}

\newcommand{\ndnname}[1]{\texttt{#1}}
\newcommand{\descr}[1]{\medskip\noindent\textbf{#1}}
\renewcommand{\paragraph}[1]{\descr{#1.}}

\begin{document}
\title{Covert Ephemeral Communication \\in Named Data Networking}

\author{
	\IEEEauthorblockN{Moreno Ambrosin, Mauro Conti}
    \IEEEauthorblockA{University of Padua\\
       Padua, Italy\\
       email: surname@math.unipd.it}
\and
	\IEEEauthorblockN{Paolo Gasti}
    \IEEEauthorblockA{New York Institute of Technology\\
    	New York, NY, USA\\
     	email: pgasti@nyit.edu}
\and
	\IEEEauthorblockN{Gene Tsudik}
	\IEEEauthorblockA{University of California\\
    	Irvine, CA, USA\\
		email: gts@ics.uci.edu}

}

\maketitle

\begin{abstract}
In the last decade, there has been a growing realization that the current Internet Protocol is 
reaching the limits of its senescence. This has prompted several research efforts that aim to 
design potential next-generation Internet architectures. Named Data Networking (NDN), an 
instantiation of the content-centric approach to networking, is one such effort. In 
contrast with IP, NDN routers maintain a significant amount of user-driven state. In this paper 
we investigate how to use this state for covert ephemeral communication (CEC). 
CEC allows two or more parties to covertly exchange ephemeral messages, i.e., messages that 
become unavailable after a certain amount of time. Our techniques rely only on network-layer, rather than application-layer, services. This makes our protocols robust, and 
communication difficult to uncover. We show that users can build high-bandwidth CECs 
exploiting features unique to NDN: in-network caches, routers' forwarding state and name 
matching rules. We assess feasibility and performance of proposed cover channels using 
a local setup and the official NDN testbed.

\end{abstract}

\section{Introduction}
\label{sec:intro}

The current IP-based Internet architecture represents an unprecedented success story, wildly  
exceeding its designers' expectations in terms of adoption, size of deployment and scalability.
Part of IP's success is due to its light-weight design: virtually all state used for 
communication is maintained at the endpoints, rather than within the network. For this reason, 
IP-based networks are -- arguably, by design -- extremely 
robust against random failures. 
However, lack of in-network state is the reason for some of IP's shortcomings, including 
poor support for efficient large-scale content distribution. 

Content distribution currently 
accounts for most Internet traffic~\cite{traffic-distribution}. 
Therefore, most major services~\cite{youtube,itunes,facebook,google}  have been -- for 
performance, cost and reliability reasons~\cite{akamai} -- relying on Content Distribution 
Networks (CDNs): 
large, complex, geographically distributed infrastructures implemented at various layers of the 
networking stack that efficiently deliver content to end users. This state of affairs 
motivated research into new  networking architectures that can better 
serve today's Internet traffic. Named Data Networking (NDN)~\cite{ndn} is 
one of these architectures. 

NDN is an example of Content-Centric Networking (CCN).
In NDN, location-agnostic content is directly addressable by name, 
regardless of who publishes it. This allows routers to store a copy of forwarded data 
in a local cache, which can be used to satisfy subsequent requests. Content is 
requested using a special kind of packets, called {\em interests}. Interests are routed 
similarly to IP packets; however, content is forwarded along the reverse path  
traversed by the corresponding interest. Data forwarding information is stored by routers, for 
a short amount of time, in a data structure called Pending Interest Table (PIT). 

User-driven soft-state on routers facilitates efficient content distribution at the network 
layer. However, availability of this state within the network creates a new set of  
problems. In particular, NDN prompts new security~\cite{ndn-dos,ersin_dos,poseidon,WahlishSV12,pollution,haining} 
and privacy~\cite{AcsCGGT13,andana} issues. In 
this paper, we investigate whether router state can be used for covert and ephemeral 
communication. We show how two parties can secretly communicate, without directly exchanging 
any packets, and without injecting new content into the network (i.e., without 
publishing new data). This is a significant departure from what can be done with IP, where lack 
of user-driven state within the network forces users to rely on the application layer for 
implementing covert channels.

We believe that this work is both timely and important. The former -- because of a recent surge 
of interest in content-centric networking, and in NDN in particular. The latter, because, to 
the best of our knowledge, it represents the first attempt to identify and address covert 
ephemeral communication (CEC) in a content-centric architecture.
CEC is, in fact, relevant in many realistic scenario, e.g.:
\begin{compactenum}
\item In tightly-controlled environments, where mandatory access control is in place (e.g., in 
the military), CEC can be used to exfiltrate sensitive information, possibly collected by  malware. Ephemeral nature of published data makes subsequent forensic analysis 
difficult.
\item In countries with oppressive governments, civil rights activists can covertly 
communicate to coordinate and exchange information. 
CEC offers plausible deniability.
\end{compactenum}

Studying whether CEC in NDN is possible -- and how to implement it -- is an important step 
towards fully understanding this means of communication, regardless of whether NDN sees limited  
deployment (e.g., as an overlay on top of IP) or widespread adoption (i.e., as a replacement 
for IP).

With this motivation, we design several protocols for exchanging covert ephemeral messages 
(CEMs) between a single 
sender and one or more receivers. We perform extensive evaluation of our techniques on a local 
network and on a geographically distributed NDN testbed. Our experiments confirm that 
CEC is indeed possible, and show that our techniques provide high 
bandwidth and low error rate.

\paragraph{Organization} We present an overview 
of NDN in Section~\ref{sec:overview}. Section~\ref{systemmodel} introduces the system model. We 
present the delay-based CEC mechanisms in Section~\ref{sec:singlebit_timingchannels} and 
common-prefix-based CEC techniques in Section~\ref{sec:namespacechannel}. 
Section~\ref{sec:error_handling} discusses sources of error and error handling. Experimental 
results are described in Section~\ref{sec:evaluation}. Security analysis is discussed in 
Section~\ref{sec:securityanalysis}. Section~\ref{sec:related_work} reviews related work. We 
conclude in Section~\ref{sec:conclusion}.

\section{NDN Overview}
\label{sec:overview}
{\em In this section we present an overview of NDN. Readers familiar with NDN may skip this section without loss of continuity. }

NDN is a networking architecture based on named data. Data is requested via {\em interests}, 
and delivered in {\em data packets}~\cite{CCNxNodeImplementation}. 
Data packets include a name, a payload and a digital signature computed by the content 
producer.\footnote{Data packets also carry additional fields that are not relevant to this 
paper and are therefore ignored.} 
A name is composed of one or more components, which have a hierarchical structure. In NDN 
notation, ``\ndnname{/}'' separates name components, e.g., \ndnname{/cnn/politics/frontpage}. 
Content is delivered to consumers only upon explicit request, which can include the full name 
of a particular data packet or a prefix of such a name -- e.g., 
\ndnname{/cnn/politics} is a prefix of \ndnname{/cnn/politics/frontpage}. In case of multiple 
data packets under a given name (or prefix), optional control information can be carried within 
the interest to restrict desired content. If no additional information is provided, producers 
and routers return arbitrary data packets matching the request (preferably, from a local 
cache).

If no local copy of a data packet is available, NDN routers forward interests towards content 
producers responsible for the requested name, using name prefixes (instead of today's IP 
address prefixes) for routing. Each NDN router maintains a Pending Interest Table (PIT) -- a 
lookup table containing outstanding [{\em interest}, {\em arrival-interfaces}] entries.
When an NDN router receives an interest, it first looks up its PIT to determine
whether another interest for the same name is currently outstanding. There
are three possible outcomes: 
(1) If the same name is already in the router's 
PIT and the arrival interface of the present interest is already in the set of 
{\em arrival-interfaces} of the corresponding PIT entry, the interest is discarded. 
(2) If a PIT entry for the same name exists, yet the arrival interface is new, the
router updates the PIT entry by adding a new interface to the set. 
The interest is not forwarded further.
(3) Otherwise, the router creates a new PIT entry and forwards the 
present interest.
We refer to (1) and (2) as PIT hit, and to (3) as PIT miss.

Upon receipt of the interest, the producer injects a matching data packet into the network, thus 
{\em satisfying} the interest. The requested content is then forwarded towards the consumer, 
traversing -- in reverse -- the path of the corresponding interest. Each router on this path 
deletes the PIT entry corresponding to the satisfied interest. In addition, each router caches a 
copy of forwarded content in its local cache.

Unlike their IP counterparts, NDN routers can forward interests out on multiple interfaces 
simultaneously. This is done in order to maximize the chances of quickly retrieving requested 
content. A router that receives an interest for already-cached content does not forward the 
interest further; it simply returns cached content and retains no state about the interest.

Not all interests result in content being returned. If an interest encounters either a router 
that cannot forward it further, or a content producer that has no such content, no error packets 
are generated. PIT entries for unsatisfied interests in intervening routers are removed after a 
predefined {\em expiration} time. The consumer can choose whether to regenerate the same 
interest after a timeout.

\section{System Model}
\label{systemmodel}

A CEC system involves a sender (\sender) and one or more receivers 
(\receiver). \sender\ wants to covertly publish a {\em time-bounded} (i.e., ephemeral) message 
$M$, while \receiver\ wants to retrieve it. A time-bounded message can only be read for a given 
period of time~\cite{CastellucciaCFK11}, after which it becomes unavailable, i.e., it {\em 
expires}. Depending on the scenario, the action of retrieving a CEM either makes it expire 
immediately, or ``refreshes'' it, hence deferring its expiration.

\snr\ are not allowed to communicate directly. For example, the Internet provider of \snr\ 
might monitor all activity between its users. Moreover, \snr\ are not allowed to use services 
(such as email or on-line forums) to exchange data indirectly. \snr\ have 
access to a producer (\producer), which is unaware of \snr's intent to communicate, and 
only hosts content that cannot be modified by consumers.

\begin{figure}[htbp]
	\centering
	\includegraphics[scale=0.5]{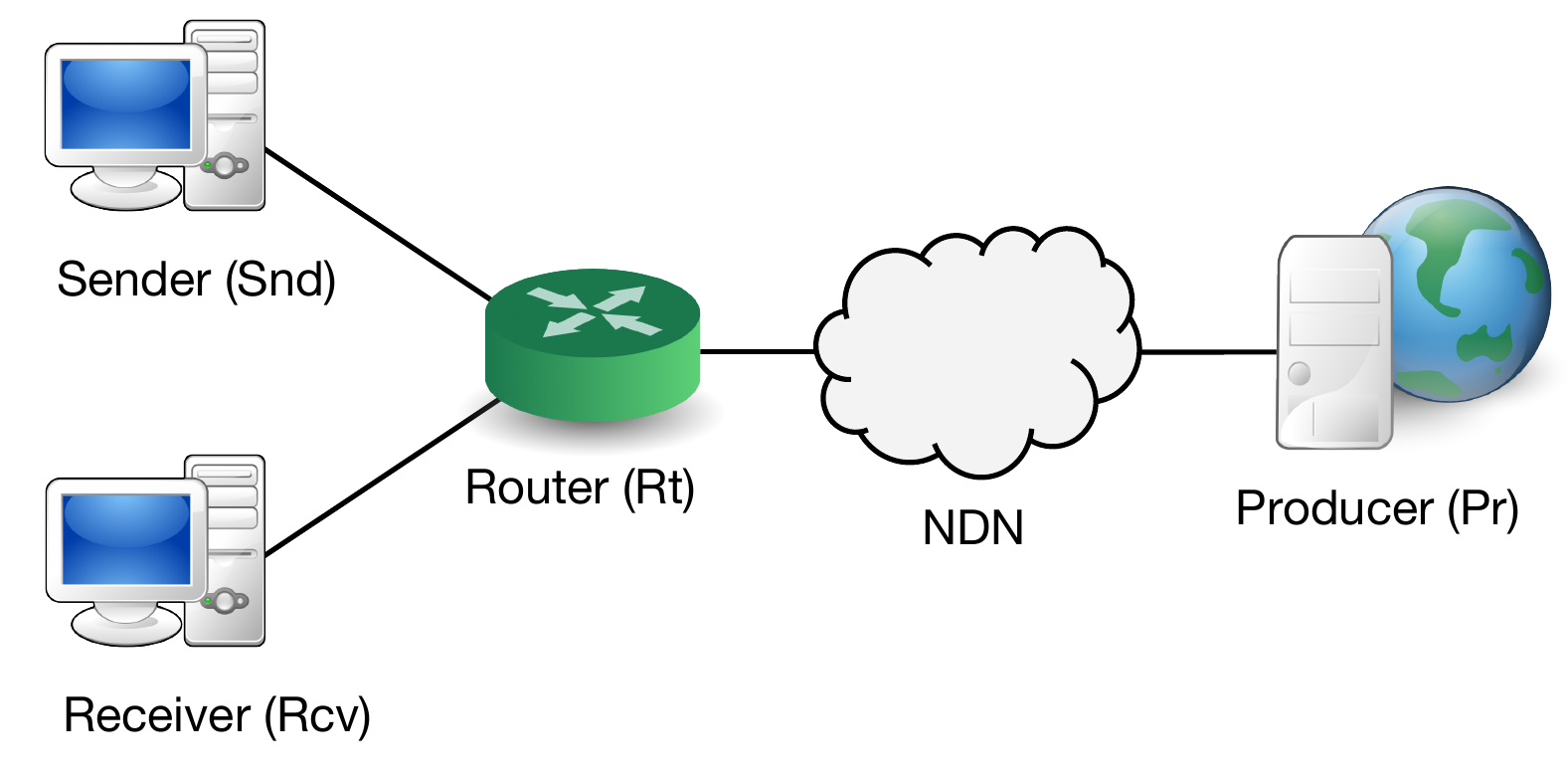}
	\caption{System model.}
	\label{img:system_model}
\end{figure}

All packets to and from \producer\ are routed through an NDN router (\router), which caches all 
data packets it forwards. At first we will assume that \router\ is \snr's first-hop router. We 
will then relax this assumption, allowing \router\ to be an arbitrary number of hops away from 
both. Figure~\ref{img:system_model} depicts our model.

We assume that \snr\ have tightly synchronized clocks.\footnote{However, this is not required in all our protocols.} 
We believe that this assumption is realistic: two parties can use NTP servers or 
GPS devices to synchronize their clocks accurately, i.e., within 500 ns to a few 
milliseconds, depending on the synchronization protocol~\cite{spectracomcorp}.

The adversary (\adversary) has three goals: (1) detecting 
CEMs from \sender\ to \receiver; (2)  preventing \snr\ from communicating; and (3) accessing 
CEMs after they expire. \adversary\ can monitor and modify traffic between users. Following 
the {\em retroactive privacy} definition of~\cite{CastellucciaCFK11}, we say that a CEC system 
is secure if any efficient \adversary\ can win the following game with probability at most 
negligibly over 1/2:

\begin{compactenum}
\item \adversary\ selects two same-length message $M_0$ and $M_1$, and sends them to \sender.
\item \sender\ selects a random bit $a$ and publishes $M_a$. 
\item After $M_a$ is expired, \adversary\ tries to retrieve $M_a$.
\item \adversary\ outputs its guess $a'$ for $a$; \adversary\ wins if $a'=a$.
\end{compactenum}

In all the proposed CECs, after \sender\ has sent a CEM, it deletes locally. 
Similarly, \receiver\ deletes all CEMs soon after receiving them, i.e., before 
the messages expire. We assume that all parties can effectively delete data.

\section{Delay-Based Covert Communication}
\label{sec:singlebit_timingchannels}

Delay-based communication relies on the ability of \receiver\ to differentiate between a cache 
(or PIT) hit, and a cache (PIT) miss. \sender\ can exploit this by selecting a set of packets 
for which it issues interests, therefore causing cache/PIT hits for \receiver.

As a warm-up, we show how timing information can be used to covertly transmit a 
single-bit CEM from \sender\ to \receiver.
Then, we describe how this can be efficiently extended to CEMs of 
arbitrary length. To simplify our notation, we refer to the RTT of a pair interest-data packet 
as the RTT of the data packet.

\subsection{Single-Bit Transmission via Cache}\label{sec:sbtc}
We now show how \sender\ sends $b\in\{0,1\}$ to \receiver. If $b=1$,  \sender\ requests 
a data packet $C$. Otherwise, it does nothing. \receiver\ determines the value of $b$ by 
requesting the same data packet $C$. If the RTT of 
$C$ is below the expected RTT for non-cached data packets, \receiver\ sets $b'$ to $1$. 
Otherwise, $b'$ is $0$.
This mechanism works reliably, i.e., $b'=b$ with overwhelming probability, if the following  
conditions are met: 
\begin{compactenum} %
\item \snr\ agree ahead of time on a data packet that will be used for communicating, and when 
\sender\ will send $b$.
\item $C$ must be non-popular, i.e., it should not be in \router's cache prior to \sender's 
request.
\item There must be separation between the RTTs associated with cache hits and cache misses, 
and \receiver\ must have a good estimate for at least one of them with respect to $C$.
\item \router\ should cache data packets for a non-negligible amount of time.
\end{compactenum}
We believe that 1 and 2 can be easily satisfied in practice.
With respect to  3, in order to distinguish a cache hit from a miss, \receiver\ must 
determine an appropriate threshold value $t_\mathit{thresh}$: iff the RTT of $C$ is below $t_
\mathit{thresh}$, then \receiver\ considers $C$ as originating from a nearby cache. 
$t_\mathit{thresh}$ can be estimated by requesting (more than once) a large number of non-
popular data packets from the same producer that distributes $C$. The first interest for each 
data packet will be satisfied by the producer itself. All subsequent (closely spaced) requests 
for the same data packet will come from a nearby cache. Regardless of the network topology, 
there is usually a clear separation between cache hits and cache misses (see 
Section~\ref{sec:evaluation}, figures~\ref{img:cache_hit_miss_distributions_LAN} 
and~\ref{img:hit_miss_distribution_testbed}) and, therefore, also an appropriate value for $t_
\mathit{thresh}$.

\receiver\ can determine if condition 4 holds by issuing multiple interests for data 
packets distributed by multiple producers, and measuring effects (if any) of content 
caching. If 4 does not hold, a different mechanism -- such as the one based on PIT -- is more 
appropriate.

We say that a CEM exchanged by \sender\ and \receiver\ is expired if $C$ has been removed 
from all caches, or once it has been retrieved by \receiver.

\begin{figure*}[t]
	\centering
	\begin{subfigure}[b]{\textwidth}\centering
		\includegraphics[scale=0.6]{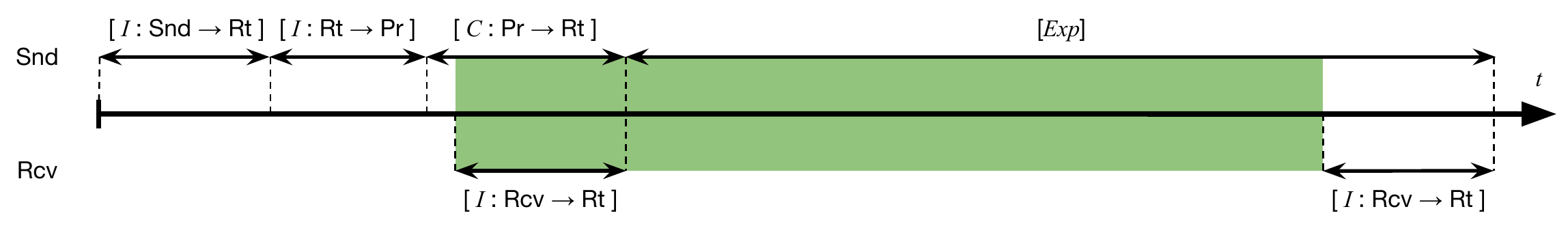}
		\caption{Cache}
		\label{img:cache_timing_channel}
	\end{subfigure}
	\\
	\begin{subfigure}[b]{\textwidth}\centering
		\includegraphics[scale=0.6]{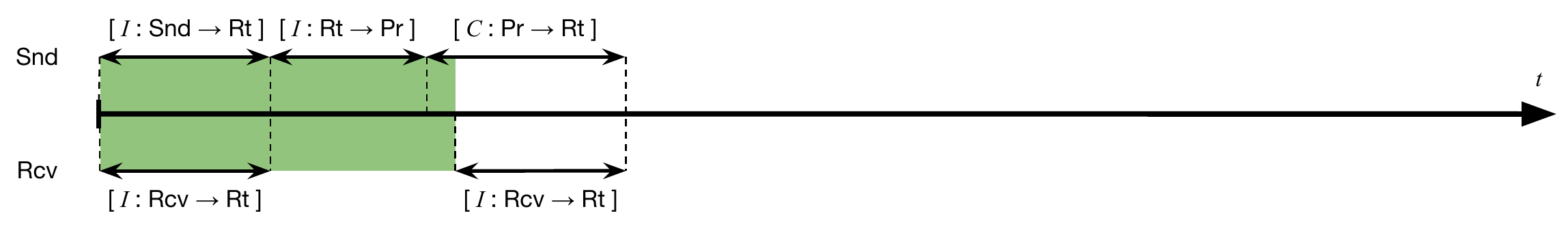}
		\caption{PIT}
		\label{img:pit_constraint_single_content}
	\end{subfigure}
	\caption{Time constrains for retrieving a CEM published using \router's cache (top) and PIT (bottom). The colored area delimits the interval in which \receiver\ can retrieve $b$.}
\end{figure*}

\paragraph{Timing Constraints}
In order to receive $b$ reliably, \receiver\ must observe a set of timing constraints. In 
particular, \receiver's interest for $C$ must be processed by \router\ after $C$ is cached (and 
made available to consumers), but before $C$ expires from the same cache. 
(Without loss of generality, in the rest of the paper we assume that data packets in \router's 
cache are available to consumers as soon as they are received by the router.) Let $I$ indicate 
an interest for $C$, and $\interest{A}{B}$, $\content{A}{B}$ the time required to 
$I$ and $C$ to travel from $A$ to $B$. Let $t_0$ be the time at which 
\sender\ writes $b$, either by issuing $I$ ($b=1$) or by doing nothing ($b=0$). Let $t_C = \interest{\sender}
{\producer} + \content{\producer}{\router}$. $C$ is available from \router's cache at $t_0 + t_C$. 
Therefore, \receiver\ can ``read'' $b$ starting at $t_b = t_0 + t_C 
- \interest{\receiver}{\router}$. When $\interest{\sender}{\router} \approx \interest{\receiver}{\router}$, $t_b 
\approx t_0 + \mbox{RTT}_{\router\rightarrow\producer}$ where $\mbox{RTT}_{\router\rightarrow
\producer}$ represents the RTT for $C$ between \router\ and \producer. \receiver\ must retrieve 
$b$ before $t_b + \expiration$, where $\expiration$ represents the freshness field of $C$, or 
the time after which $C$ is evicted from \router's cache, whichever comes first. 
Figure~\ref{img:cache_timing_channel} summarizes these observations.

Time needed to read a single bit depends on the RTT associated with a cache hit, from 
\receiver's point of view. Let $\hit$ and $\miss$ indicate the average RTT for a cache hit and 
cache miss relative to $C$, as observed by \receiver. 
\receiver\ sets $b=1$ iff the RTT of $C$ is below $\hit + \Delta<\miss$, 
where $\Delta$ is a small constant used to account for variance in $C$'s RTT. \router\ can 
therefore determine $b$ within $\hit + \Delta$.

Covert messages distributed with this technique are ephemeral, i.e., they become unavailable 
after a certain amount of time without any further action from \sender\ or \receiver. Because 
\router\ caches forwarded traffic, $C$ will eventually be evicted from 
\router's cache. In fact, we argue that $C$ is always a good candidate for deletion: since $C$ 
is not popular, both Least Frequently Used (LFU) and Least Recently Used (LRU) cache replacement 
policies will consider it for removal relatively early. 

Once \receiver\ requests $C$, it will be stored in cache regardless of the original value of 
$b$. Therefore, after being retrieved, $b$ will be set to 1 until $C$ is evicted from \router's 
cache.

Our experiments, reported in Section~\ref{sec:evaluation}, show that this technique provides 
high bandwidth, with low error. Moreover, it is relatively easy to 
implement, since it does not require strict time synchronization.

\subsection{Single-Bit Transmission via PIT}\label{sec:sbtp}
In some circumstances, cache-based %
CEC 
	is not applicable:
\begin{compactenum}
\item \router\ might have no cache: small, low-cost, low-power embedded routers may not store 
forwarded data packets. 
\item \router's entire cache may be overwritten before \receiver\ 
issues $I$. This can happen if \router's cache is very small, and the router forwards a large 
amount of traffic.
\item To prevent cache pollution attacks~\cite{pollution,haining}, \router\ may not add to 
its cache data packets that are forwarded only once. This behavior would force \sender\ to issue 
$I$ multiple times before $C$ is stored in \router's cache, negatively affecting both 
bandwidth and detectability.
\item \router\ may implement cache privacy techniques that involve delaying serving $C$ when it 
is retrieved from the cache~\cite{AcsCGGT13}.
\end{compactenum}

To overcome the above limitations, we design a  technique based on PIT state. This technique 
requires strict time synchronization between parties. It is based on PIT hits 
(see Section~\ref{sec:overview}): when \router\ receives 
interest $I'=I$, while $I$ is still in \router's 
PIT, the two interests are ``collapsed'' within the same PIT entry. \router\ adds 
the incoming interface of $I'$ to the PIT entry of $I$, and does not propagate $I'$ any further. 
Once $C$ is received by \router, it is forwarded to the interfaces on which $I$ and $I'$ 
were received.

This feature of NDN can be used by \snr\ to covertly exchange a bit $b$ as follows. If $b=1$, 
\sender\ issues $I$, otherwise it does nothing. To receive $b$, \receiver\ issues $I'=I$ while 
a copy of $I$ intents is still in \router's PIT -- if originally issued by \sender. If $I$ 
is in \router's PIT, then \sender's interest will be satisfied more quickly than if it was not, 
because either the original $I$ would be already past \router, or $C$ would be 
already on its way back to \sender. If \receiver\ can correctly measure the corresponding 
difference in RTT, it can reliably receive $b$.

With this technique, we say that a CEM has expired if $I$ has been removed from \router's PIT 
{\em and} from all caches, or it has been retrieved by \receiver.

\paragraph{Timing Constraints}
While the PIT-based CEC works regardless of \router's cache behavior (or even cache 
availability),  
it imposes much stricter timing requirements on \receiver. In fact, $I'$ must be received by \router\ after $I$ (if 
issued) is added to \router's 
PIT. Moreover, $I'$ must be received before $C$ is forwarded to by \router. This gives to 
\receiver\ a time window of $\mbox{RTT}_{\router\rightarrow\producer}$.

As in the cache-based technique, messages exchanged via PIT are ephemeral: if $I$ is not 
issued on time, the corresponding PIT entry will be removed once $C$ is forwarded back to \sender. Also, 
after \receiver\ issues $I$, any attempt to retrieve $b$ within the correct timing constraints 
will result in a collapsed interest (and therefore set $b=1$), regardless of the original value 
of $b$.
Figure~\ref{img:pit_constraint_single_content} gives a graphical representation these constraints.

\subsection{Tandem Data Packets}\label{sec:tdp}
With geographically distributed deployments of NDN, and when \router\ is far 
from \receiver, RTT associated with cache hits and misses may fluctuate significantly over 
time. In order to reduce the probability of erroneously detecting cache hits/misses, we introduce 
a technique -- called Tandem Data Packets (TDP) -- that uses two data packets to covertly receive a single bit. To transmit $b$, \snr\ 
agree on data packets $C_0$ and $C_1$, which are assumed not to be in any cache.  Then, 
\sender\ requests $C_b$. \receiver\ issue two consecutive interests, one for $C_0$ and one for 
$C_1$; if RTT of $C_0$ is lower than RTT of $C_1$, \receiver\ sets $b'=0$, otherwise 
it sets $b'=1$. The CEM is exchanged correctly if $b'=b$.

This technique does not require \receiver\ to make any {\em a priori} assumption on the 
exact RTT associated with cache hits and misses, besides the fact that the RTT of $C_b$ is 
lower than the RTT of $C_{\neg b}$. As our experiments confirm, this reduces receiver error 
when, since RTT for both hits and misses is continuously updated according to network 
conditions. 
With this technique, after it is obtained by \receiver\ $b$ becomes inaccessible. In fact, both 
$C_0$ and $C_1$ will be stored in \router's cache, due to \receiver's interests. Therefore, any 
difference in the RTT associated with $C_0$ and $C_1$ will not depend on $b$. Therefore, $b$ 
expires if it has been removed from \router's cache or it has been retrieved by 
\receiver.

\paragraph{Timing Constraints} Timing constraints are identical to those in Section~\ref{sec:sbtc}.

\subsection{Transmission of Multi-Bit Messages}
\label{sec:multi-bit}

\snr\ may want to exchange messages composed of more than one bit. 
We discuss how to determine \sender's and \receiver's speed separately, since the two may send 
and receive at different rates.

Let $M = b_1,\ldots,b_n$ be an n-bit string. Suppose that \snr\ agree on $n$ 
different data packets $C_1,...,C_n$.
Instead of waiting for the full RTT of $C$, \sender\ can send new $I_i$ for $C_i$ before 
$C_{i-1}$ has been received. \sender\ selects an interval $t$; two consecutive interests $I_i, 
I_{i+1}$ are sent at $t_i$ and $t_{i+1}$, where $t_{i+1} = t_i + t$. The minimum value for $t$ 
is denoted as $t_\mathit{min}$, and corresponds to sending an uninterrupted burst of interests. 

Similarly, \receiver\ selects a value $t$ which is used to determine how subsequent interests are spaced. \snr\ can select $t$ independently, as long as the timing constraints associated with the protocol are not violated.

We evaluate how this technique affects transmission error as a function of $t$ and report our findings in Section~\ref{sec:evaluation}. 

\paragraph{Transmitting Multiple Bits with a Single Interest} For efficiency reasons, \sender\ 
can use a generalization of the TDP technique to send multiple bits using a single interest. 
Two parties agree a priori on a set of data packets, which we represent as a matrix: 
$$ Y = \begin{bmatrix}
        C_{(1,1)} & \cdots &  C_{(1,2^m)}           \\[0.4em]
       \vdots &   & \\[0.4em]
        C_{(\ell,1)} &  &  C_{(\ell,2^m)}
     \end{bmatrix}
$$
where $m$ is the number of bits transmitted using one interest, and 
$\ell = {\lceil n/m \rceil}$. In order to 
publish $M$, \sender\ splits it in words $W_1, \ldots, W_\ell$ of $m$ 
bits each (i.e., $W_1 = (b_1,\ldots,b_m)$, $W_2 = (b_{m+1}, \ldots, b_{2m})$, etc.). \receiver\ 
then issues interests for $C_{(1,W_1)}, C_{(2,W_2)}, \ldots, C_{(\ell, W_\ell)}$, where $W_i$ 
is used as integer representation of the corresponding bit string.
Thus, \sender\ can publish an $n$-bit message using ${\lceil n/m \rceil}$ 
interests.

To retrieve $M$, \receiver\ issues interests for all data packets in $Y$. Let 
$C_{i,j}$ be the data packet on the $i$-th row of $Y$ such that the RTT of $C_{i,j}$ is the 
smallest across all $C_{i,1}, \ldots, C_{i,2^m}$. \receiver\  sets $W_i = j$, and $M = W_1|
\ldots|W_\ell$.
The cost of retrieving $M$ for \receiver\ is therefore exponential in $m$. (In practice, 
reasonable values for $m$ are between $1$ and $5$).
Note that when $m=1$, this technique corresponds to TDP.

\section{Common-Prefix-Based Covert Communication}
\label{sec:namespacechannel}

Using previous techniques, a covert message can be retrieved only by a single receiver. Message 
is automatically ``deleted'' after it is ``read'' by \receiver. This is 
desirable when a CEM has only one intended recipient. However, when the CEM has multiple 
recipients, \sender\ must create a separate ``instance'' of the message for each. In this 
section, we propose a technique -- called Common-Prefix-Based Covert Communication (CPC) -- 
that allows \sender\ to publish a message once, and have multiple parties to retrieve it. 
Similarly to previous techniques, CEMs published using CPC are ephemeral.

CPC relies on NDN's longest prefix matching feature, instead of RTT measurements. This 
makes it robust against cache privacy techniques~\cite{AcsCGGT13}, which could 
defeat CEC techniques introduced in Section~\ref{sec:singlebit_timingchannels}.

Communication via CPC works as follows. \snr\ agree on two data packets $C_0, C_1$ which share 
a common name prefix, e.g., \ndnname{/common/prefix/C0}, and \ndnname{/common/prefix/C1}. 
\footnote{Common prefix can be followed by 
different children namespaces, e.g., 
\ndnname{/common/prefix\textbf{/foo/}C0}  and  \ndnname{/common/prefix\textbf{/yet}}
\ndnname{\textbf{/another/prefix/}C1}.}
The common namespace is selected such that data packets published under   
it are not popular, i.e., not in \router's cache. 
In order to transmit a single bit, \sender\ simply requests $C_b$. 
To receive $b$, \receiver\ issues an interest for \ndnname{/common/prefix/}. Both $C_0$ and 
$C_1$ match \receiver's interest. Therefore, \router\ will return one data packet among $C_0$ 
and $C_1$ that is still in its cache -- or in its PIT, if \snr's interests are closely spaced 
(see timing constraints below).
This communicates $b$ to \receiver.

This technique is very robust against changing network conditions. In particular, since timing 
is not used to either set or determine $b$, transient changes in RTT do not 
introduce communication errors: \receiver\ receives only $C_b$, regardless of how long it 
waits. 
Moreover, in contrast with previous techniques, when \receiver's interest is dropped (or, 
similarly $C_b$ in response to \receiver's interest is dropped) \receiver\ can re-issue 
its interest, since this process does not affect $C_b$. 

Common-prefix-based covert channels are suitable for distributing a single message to a (possibly 
large) set of receivers. Each interest for \ndnname{/common/prefix/} issued by a recipient 
has the side-effect of ``refreshing'' $C_b$ in \router's cache, making $b$ available 
longer. After recipients stop retrieving $C_b$, it ``fades away'' from all 
involved routers' caches, effectively erasing $b$.
As an alternative, \sender\ or one of the recipients can request $C_{\neg b}$ which achieves a 
similar result. 

A message exchanged using CPC expires when it is removed from all caches.

\paragraph{Timing Constraints} In order to successfully retrieve $b$, \receiver\ must issue an 
interest for \ndnname{/common/prefix/}  such that the interest is received by 
\router\ after the interest for $C_b$ from \sender. If the interest from \receiver\ is received 
before $C_b$ is returned to \receiver, communication between \snr\ is implemented through 
\router's PIT. Otherwise, \router's cache is used to exchange $b$. 
\sender's interest must also be received by \router\ before $C_b$ is removed from the cache.

\subsection{Multiple-Bit Transmission}

Since this technique is less susceptible to RTT fluctuations and packet loss, using it for 
sending and receiving multiple bits in bursts does not introduce significant errors. This is 
confirmed by our experiments, in Section~\ref{sec:evaluation}.

\paragraph{Transmitting Multiple Bits with a Single Interest} \snr\ can agree on data packets 
in matrix $Y$ with the additional requirements 
that for $i\in [1,\ell]$, data packets in row $i$ share the same common prefix 
$\mathit{pref}_i$. 
\sender\ splits $M$ in $W_1, \ldots, W_\ell$, and -- for each $i$ -- issues one interest for 
$C_{i,W_i}$.

\receiver\ needs to issue only {\em one} interest per word (i.e., per 
matrix row), requesting a data packet from $\mathit{pref}_i$. For this reason, \snr\ can 
exchange an $n$-bit message using ${\lceil n/m \rceil}$ interests/data packets each.

In practice, $m$ is limited only by availability of un-popular namespaces containing a 
sufficient number of data packets.

\section{Errors and Error Handling}
\label{sec:error_handling}

Bit errors may be introduced by both \sender\ (write errors) and \receiver\ (read errors). 
Depending on the technique used to communicate, errors may be injected in $M$ for different 
reasons and may be detected and dealt with in different ways.
A write error occurs when a data packet requested by \sender\ is not added to \router's 
cache or PIT. A read errors occurs as a result of an incorrect retrieval of a message bit after 
it has been correctly written, and before it is expired.

\paragraph{Delay-Based: Cache}
We consider the following two issues as common causes for write errors: 
\begin{compactenum}
\item \underline{Packet loss (either interests or data packets)}. Interests from \sender\ may be 
dropped along their way to \producer.  Similarly, data packets from \producer\ may be 
dropped before they reach \router. In both cases, no data packets added to \router's 
cache, and therefore the send operation fails. This, however, can be detected by \sender, who 
simply re-issues interests for which it does not receive data packets.
\item \underline{Forwarded data packets not added to \router's cache}. This can 
be caused, for example, by meta-cache algorithms on \router. \sender\ can detect this only 
by re-requesting all bits set to 1 in its messages and, for each comparing the RTT of the 
first request with the RTT of the second.
\end{compactenum}
We identify the following causes for read errors: 
\begin{compactenum}
\item \underline{RTT fluctuations}. Since retrieving a message relies on correctly identifying cache 
hits and misses, any overlap in the RTT between \receiver\ and \router\ and between \receiver\ 
and \producer\ could cause a read error. These errors are not detectable, and cannot be 
addressed by simply re-sending interests.
\item \underline{Interests from other consumers}. Some consumers may request a data packet 
that correspond to a bit in the message set to 0, and have it added to \router's cache. We 
assume that this happens with negligible probability, since \snr\ exchange messages using a set 
of data packets that are not popular.
\item \underline{Packet loss (data packets)}. If a data packet is dropped on the path 
from \producer\ to \router, it can be safely be re-requested by \receiver\ without altering the 
original message. However, if it is dropped on its way from \router\ to \receiver, the 
corresponding message bit will be set to 1 regardless of its original value. \receiver\ can only distinguish between the two cases -- and determine the correct value of the corresponding message bit $b$ -- when $b$ is read as $0$.
\item \underline{Packet loss (interests)}. When interests are dropped on their way from \receiver\ to 
\router\ (if the corresponding data packet is in \router's cache) or to \producer\ (if it is 
not), \receiver\ cannot retrieve the corresponding bit. In this case, \receiver\ can re-issue 
the same interest without altering the original message, since no data packets have been 
added to \router's cache. However, since loss of interest cannot be distinguished from loss of data packet, \receiver\ may not be able to recover from this error.
\item \underline{\router\ is rebooted.} This causes all data packets in \router's cache to be 
deleted, therefore ``erasing'' all messages from \sender. This can be detected if \receiver\ knows that $M\neq0^n$.

\end{compactenum}

\receiver\ can reduce errors induced by RTT fluctuations using the ``scope'' field in 
interests, when \router\ is its first-hop router. This field works similarly to the IP TTL 
field. When scope is set to 2, interests are forwarded for up to one hop. (Values higher than 2 
are not allowed~\cite{CCNxInterests}). If the \receiver's first hop cannot satisfy the 
interests, it simply drops it. This way, \receiver\ does not need to measure any difference in 
the delay of cache hits and misses, since only cache hits will result in returned content. 
Moreover, this would allow interest retransmission in case of packet loss, since setting scope 
to 2 would prevent \receiver's interests from adding any new content into the cache. We argue 
that, however, setting the scope field would make \receiver's activity easier to detect.

\paragraph{Delay-Based: PIT} As in to the previous technique, write errors 
correspond to interests sent by \sender\ and are not added to \router's PIT.
The main cause for write errors is loss of the interest from \sender\ to \router. This cannot 
be detected on time by \sender, since the same interest must be issued by \receiver\ 
before the corresponding data packet is received by \sender.

On the receiver side, errors may have the following causes:
\begin{compactenum}
\item \underline{RTT fluctuations.} Similarly to the previous technique, significant fluctuations of 
RTT can introduce read errors. 
\item \underline{Packet loss (either interests or data packets)}. In case of packet loss, 
\receiver\ will learn no information about the corresponding bit in the covert message. 
Moreover, re-transmitting an interest may provide no useful information, since by then the PIT 
entry corresponding to the original interest from \sender, if any, will be either expired or 
removed.
\item \underline{Interests from other consumers.} Other consumers may issue the same interests that 
\snr\ are using to covertly exchange information. However, this happens with negligible 
probability, because: (1) data packets used to covertly publish messages are non-popular, and (2) 
interests from other consumers must be issues a few milliseconds before \receiver\ issues its 
interests.
\item \underline{Lack of synchronization between \snr}. Depending on the topology, \snr\ must be 
tightly synchronized, i.e., roughly within half RTT between \sender\ and \producer. Lack of 
synchronization may lead to a high rate of read errors.
\item \underline{Message expiration}. Even though this technically is not a read error, it may happen 
that \receiver\ cannot retrieve part of the message on time due to the strict timing 
requirements.
\end{compactenum}

As before, the scope field can be set in \receiver's interest to 
reduce error rate.

\paragraph{TDP} Write errors have the same 
causes, as well as detectability, as the write errors in delay-based cache technique. 

Similarly, read errors have the same causes as with delay-based, single-bit cache. However, 
data packet-pairs provide more robustness against RTT fluctuations and packet loss. Since two subsequent RTTs -- one 
corresponding to a cache hit, and one for a cache miss -- are measured for each message bit, the 
probability of error associated with random RTT fluctuations is greatly reduced.
With respect to packet loss, at least one of the data packets corresponding to a single 
message bit will be returned with relatively high probability. The associated RTT will still 
allow \receiver\ to estimate whether it is coming from \router's cache -- although less 
accurately. 

\paragraph{Common-prefix-based Covert Communication} Using this technique, write errors may be 
introduced by the same events that trigger packet loss in delay-based, single bit cache. 
With respect to read errors, this technique is significantly more robust than the previous ones 
because: (1) it does not rely on timing measurements, and is therefore immune to RTT  
fluctuations; and (2) in case of packet loss (affecting either interests or data packets), \receiver\ can simply re-issue its interest, without affecting the covert message.

\subsection{Error Correction} To address potential read/write errors, \sender\ can use 
error-correction codes with CEM. 
For example, Reed-Solomon error correction codes~\cite{reed-solomon} could be used. We do not 
investigate this any further, since the goal of this paper is to assess feasibility of the 
channel and the corresponding error rate.

\section{Evaluation}
\label{sec:evaluation}
We implemented a prototype CEC system to evaluate our protocols. In this section we present the 
results of our experiments. 
The prototype is based on CCNx \cite{ccnx}, an open-source implementation of NDN which runs as 
an overlay on top of IP. 
We performed experiments on the two topologies:
\begin{compactitem} 
\item \underline{{LAN}}, composed of \sender, \receiver, \router\ and \producer\ within 
the same broadcast domain. Each party runs a separate instance of CCNx.
\item \underline{NDN testbed}~\cite{ndn-testbed}, where \sender\ and \receiver\ (located in 
Europe) 
are connected to the UCLA NDN hub (which acts as \router), and \producer\ is connected to the 
testbed through the UCI hub. UCLA and UCI hubs are one NDN hop apart (ten hops over IP).
\end{compactitem}

\snr\ exchange 1,000-bit messages. Each message is a fresh random bit 
string. This is representative of the distribution of encrypted messages. 

Naturally, our protocols generate communication overhead. We used 41-byte 
interests and 377-byte data packets (on average). With single-bit transmission (either using 
PIT and cache), each message bit set to 1 requires \sender\ to exchange 418 bytes. Regardless 
of message content, \receiver\ needs to send/receive 418 bytes per message bit. With the TDP 
protocol, each message bit costs 418 bytes to \sender\  and 836 bytes to \receiver. When transmitting 
multiple bits with a single interest, $m$ message bits cost \sender\ 418 bytes, and $2^m\cdot\ $418 
bytes to \receiver. Finally, with CPC both \snr\ exchange 418 bytes for each $m$-bit word.

\subsection{Evaluation of Delay-Based (Cache) Techniques} 
\label{sec:eval-cache-hits}
In order to assess feasibility of cache-based techniques, we compared RTT associated 
with cache hits and cache misses in both LAN and testbed scenarios. 

\begin{figure}[ht]
			\begin{subfigure}[b]{0.48\textwidth}
		            \centering
		            \includegraphics[scale=0.25]{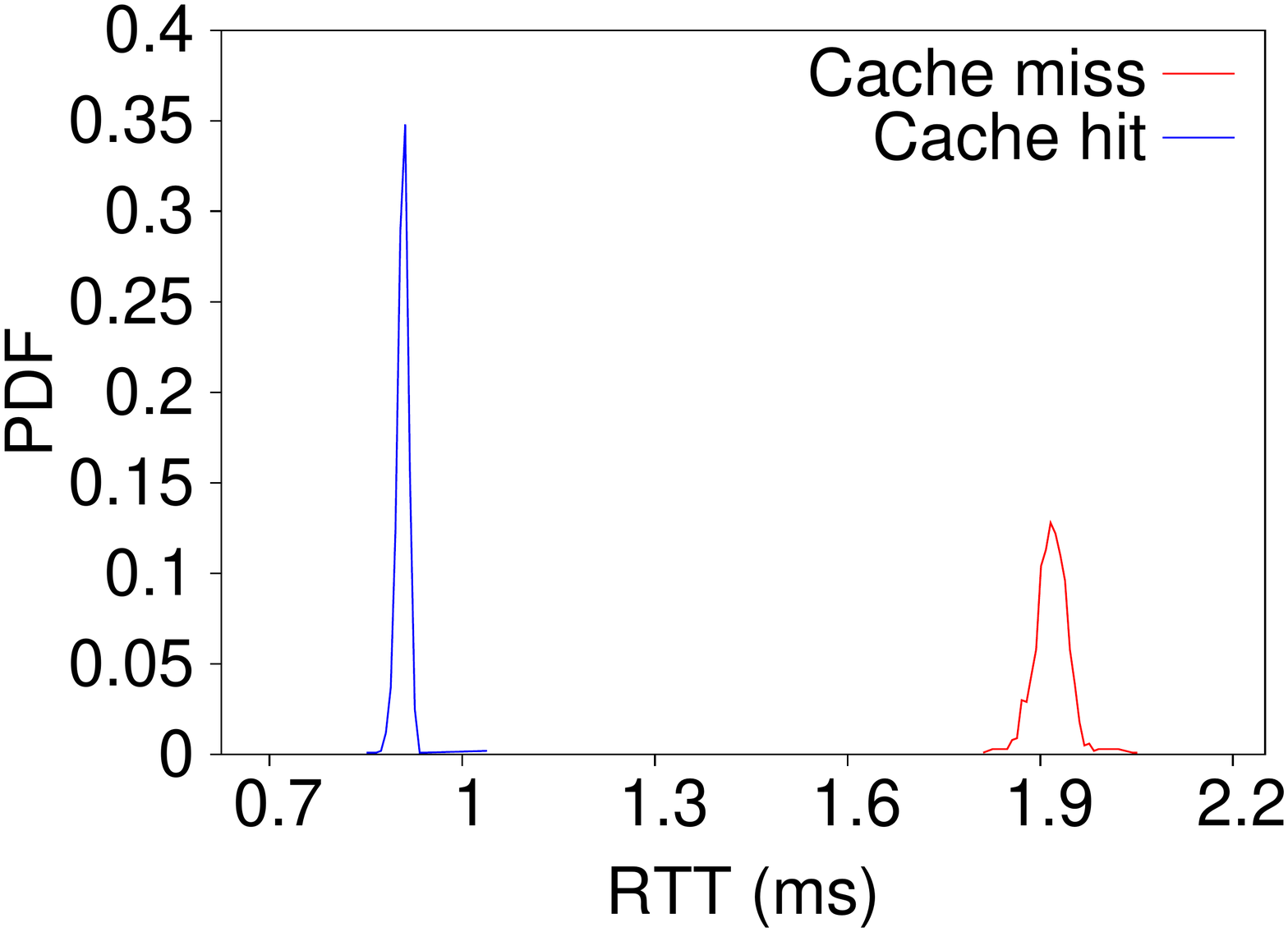}
		            \caption{LAN}
		            \label{img:cache_hit_miss_distributions_LAN}
        		\end{subfigure}
			
			\vspace{-0.4cm}
			
        		\begin{subfigure}[b]{0.48\textwidth}
		            \centering
		            \includegraphics[scale=0.25]{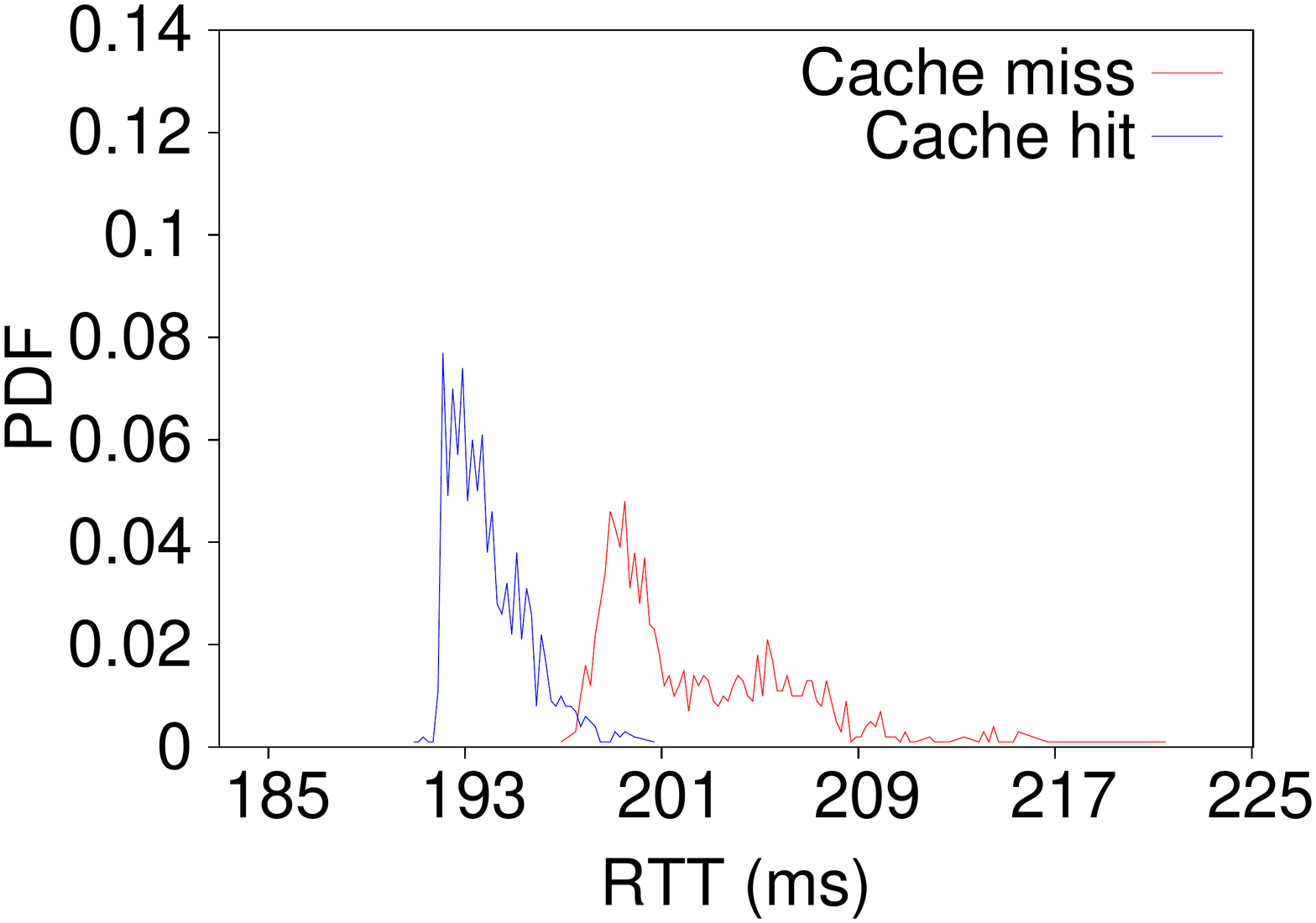}
		            \caption{Testbed}
		            \label{img:hit_miss_distribution_testbed}
        		\end{subfigure}
        		\caption {PDF for cache hit and cache miss.\label{img:hit_miss_distribution}}
	\end{figure}

Figure~\ref{img:hit_miss_distribution} summarizes our findings and represents average values 
over 100,000 data packets. While there is virtually no overlap between RTT of cache hits and 
misses in a controlled (LAN) environment, RTT fluctuations on the testbed do not always allow 
us to 
distinguish a cache hit from a cache miss. However, the overlap is still relatively small 
and, as confirmed by further experiments, it is possible to implement a reliable 
CEC on the testbed.

\begin{figure*}[t]
			\begin{subfigure}[b]{0.2\textwidth}
		            \includegraphics[scale=0.2]{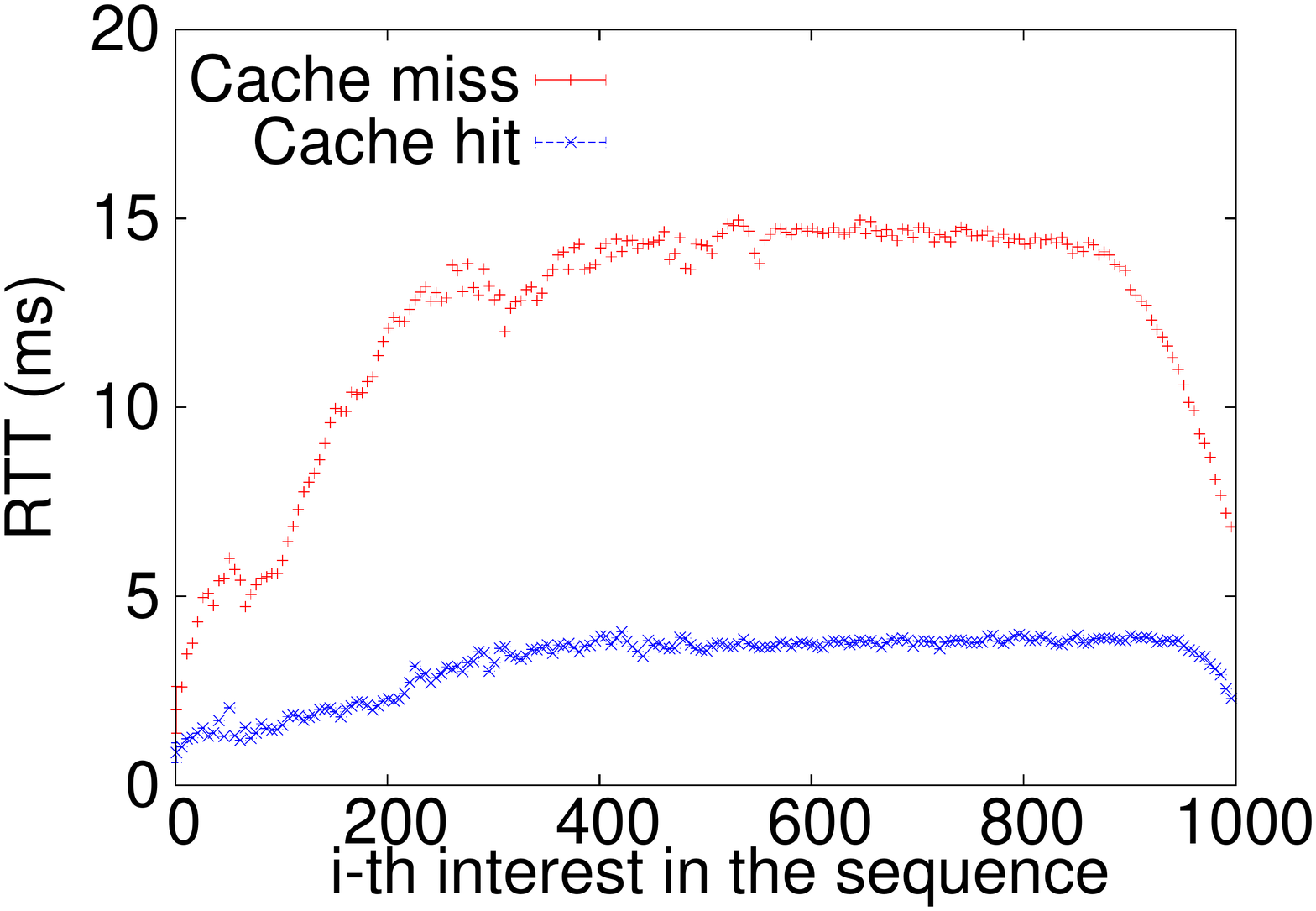}
		            \caption{LAN, $t=t_\mathit{min}$}
		            \label{img:cache_hit_miss-0.0_95_LAN}
        		\end{subfigure}
			\hspace{2cm}
			\begin{subfigure}[b]{0.2\textwidth}
		            \includegraphics[scale=0.2]{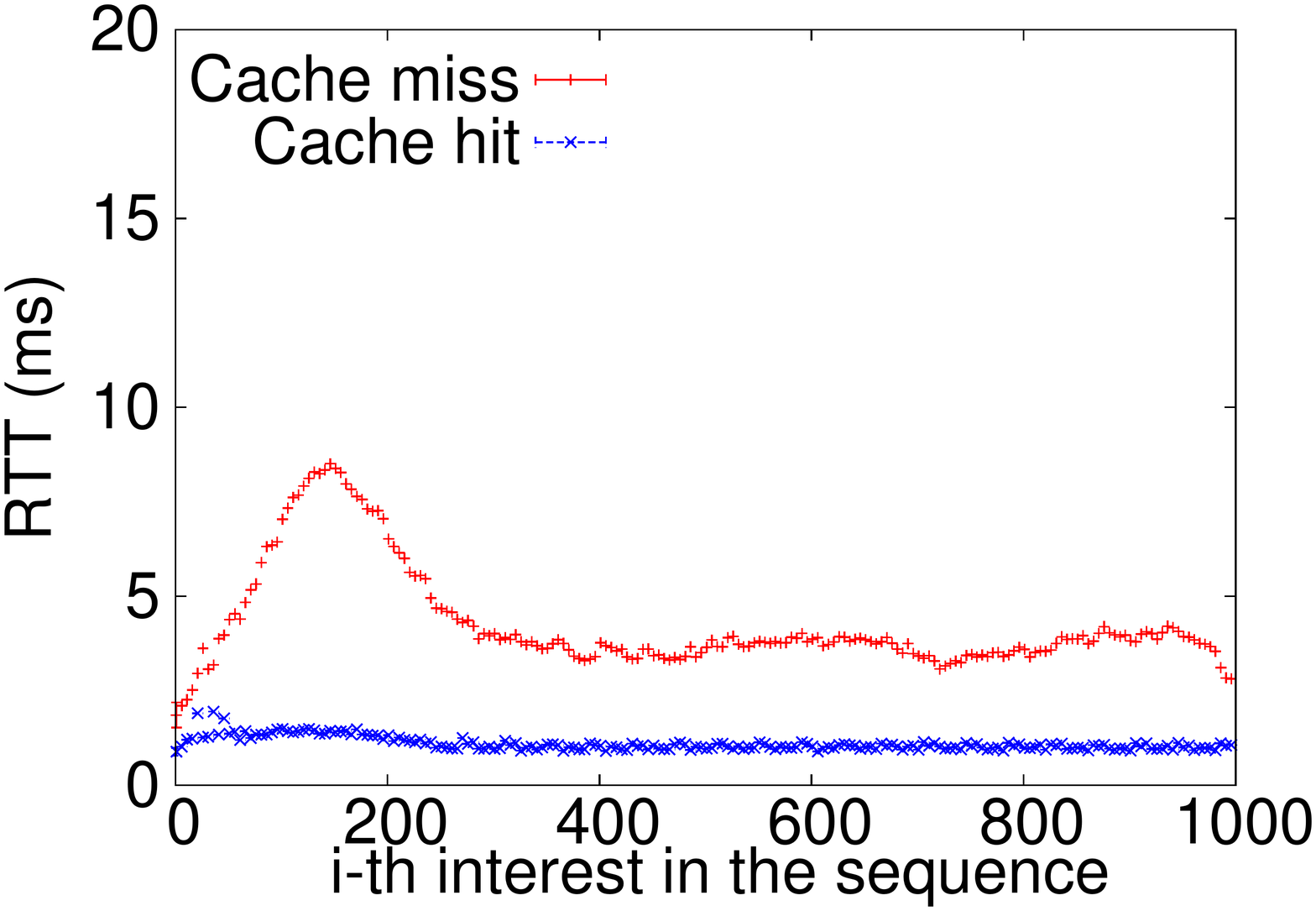}
		            \caption{LAN, $t=0.2$ ms}
		            \label{img:cache_hit_miss-0.2_95_LAN}
        		\end{subfigure}
			\hspace{2cm}
			\begin{subfigure}[b]{0.2\textwidth}
		            \includegraphics[scale=0.2]{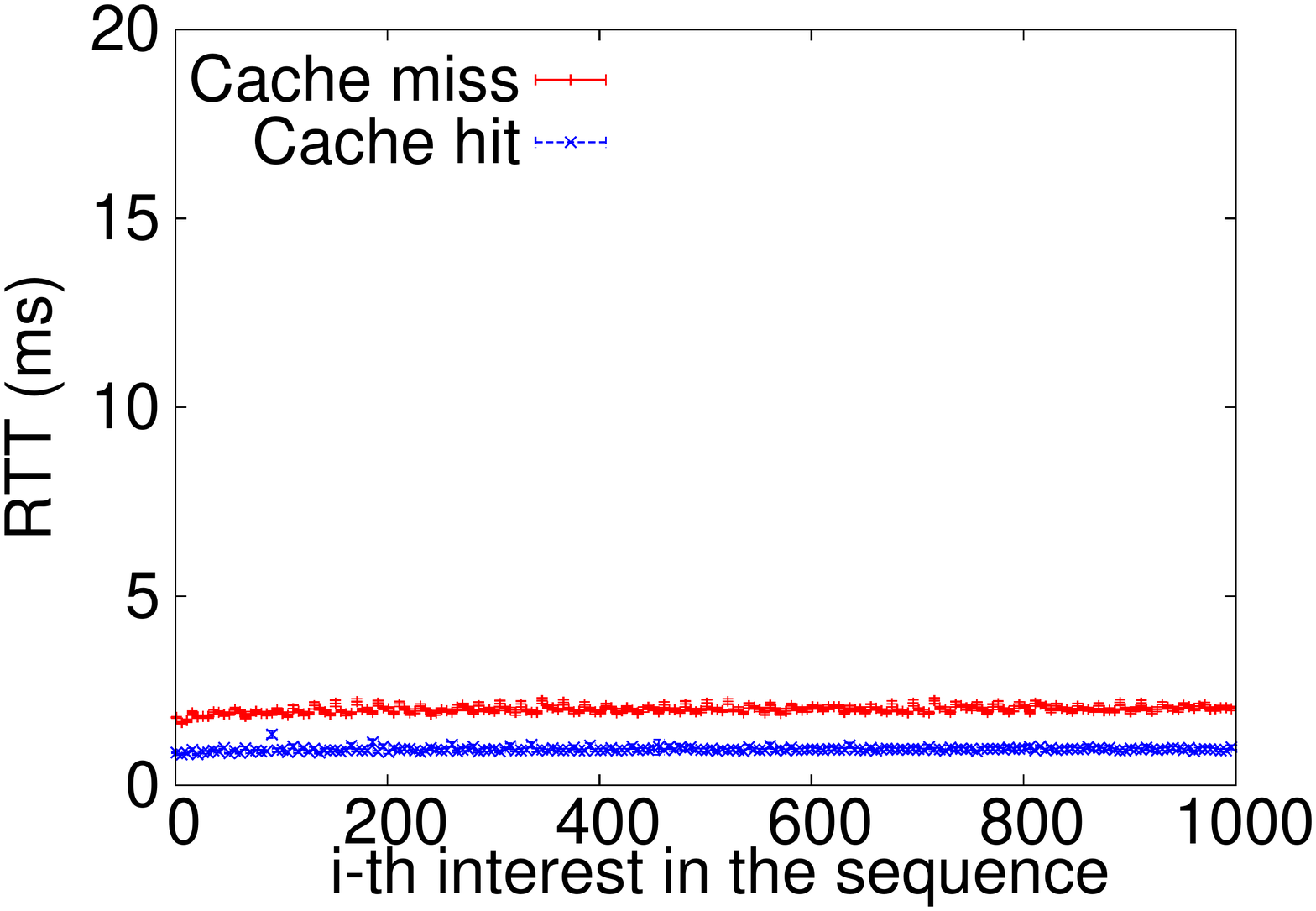}
		            \caption{LAN, $t=0.5$ ms}
		            \label{img:cache_hit_miss-0.5_95_LAN}
        		\end{subfigure}		
\newline
			\begin{subfigure}[b]{0.2\textwidth}
		            \includegraphics[scale=0.2]{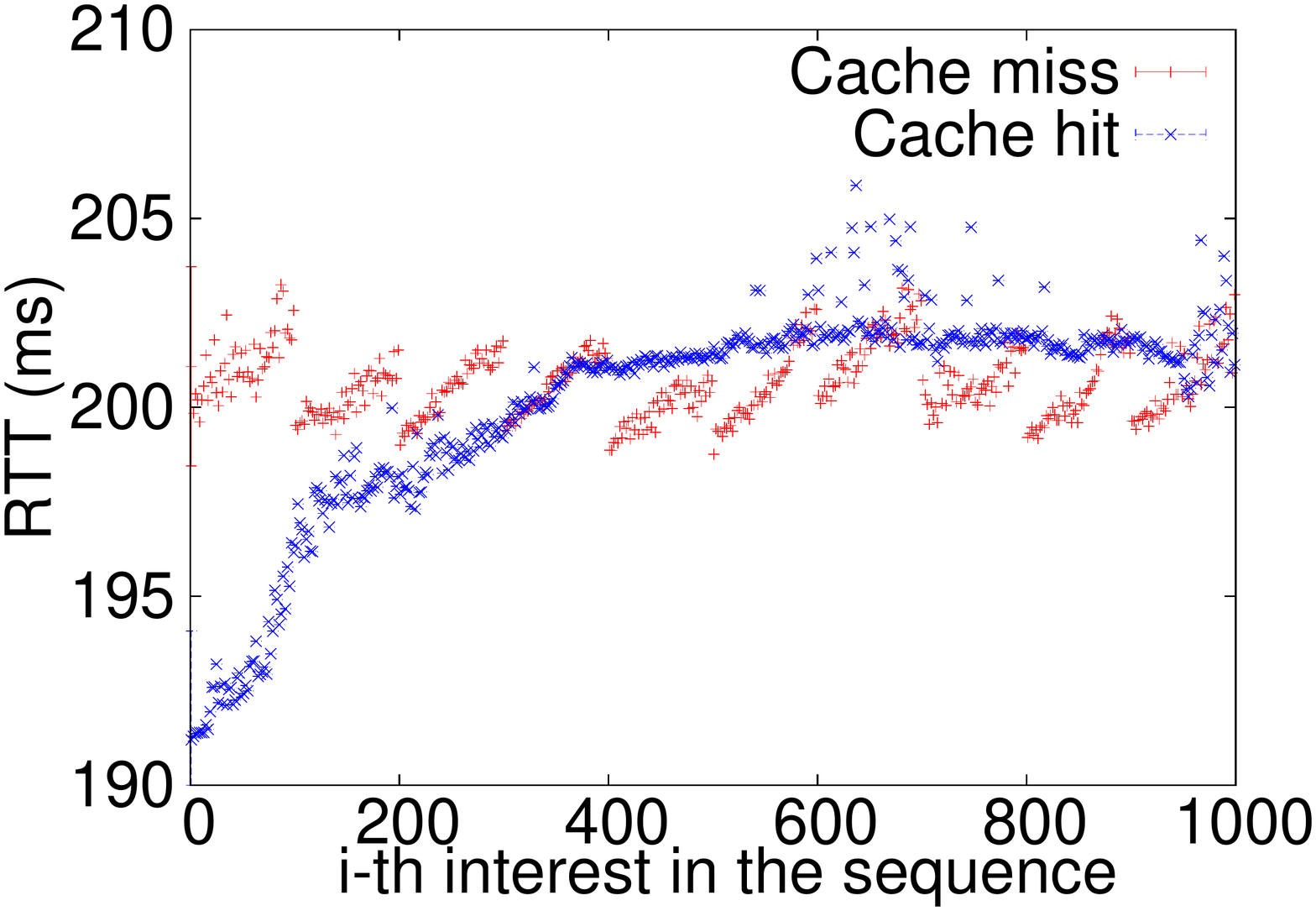}
		            \caption{Testbed, $t=t_\mathit{min}$}
		            \label{img:cache_hit_miss-0.0_95_TB}
        		\end{subfigure}\hspace{2.15cm}
        		\begin{subfigure}[b]{0.2\textwidth}
		            \includegraphics[scale=0.2]{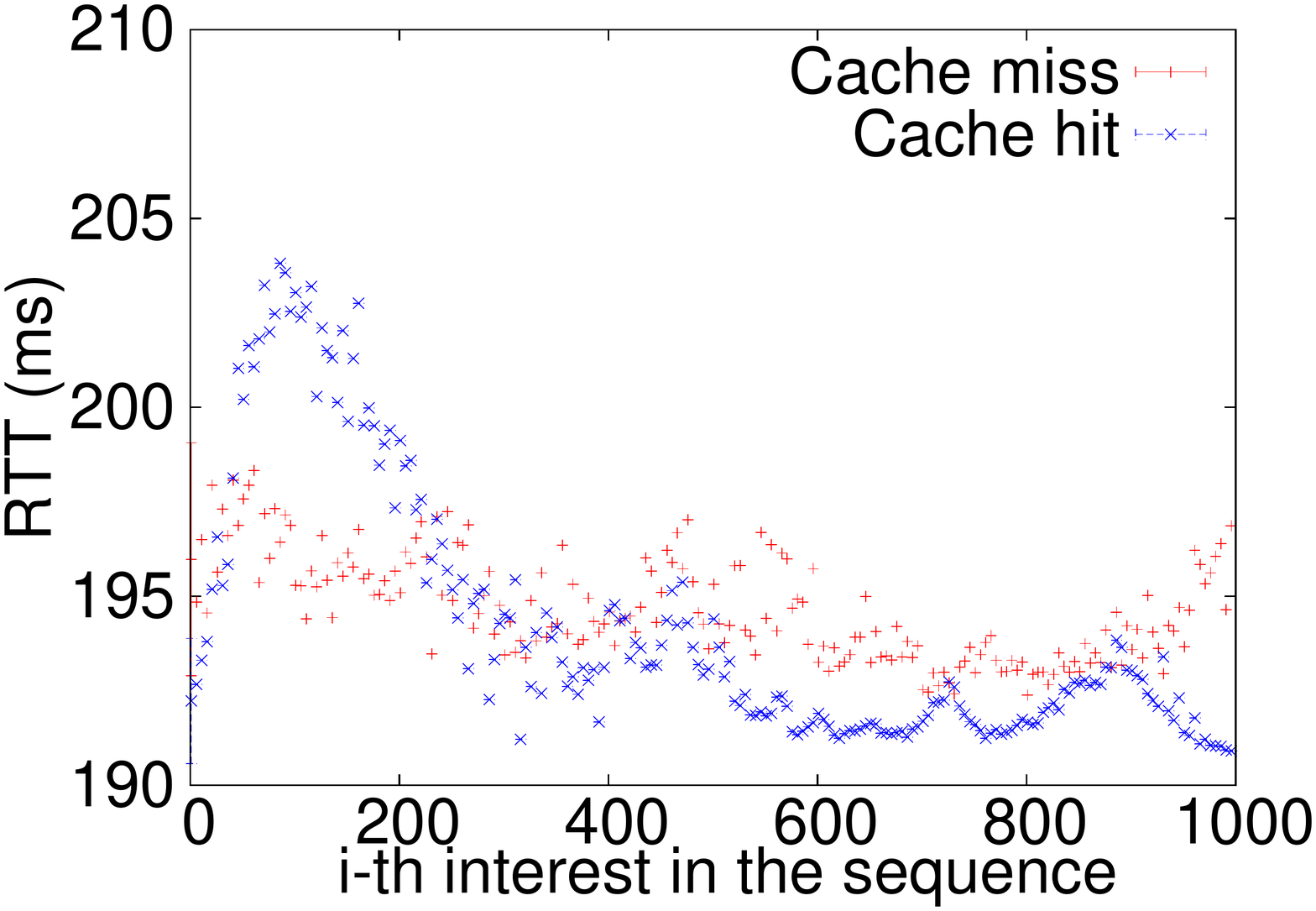}
		            \caption{Testbed, $t=0.2$ ms}
		            \label{img:cache_hit_miss-0.2_95_TB}
        		\end{subfigure}\hspace{2.15cm}
        		\begin{subfigure}[b]{0.2\textwidth}
		            \includegraphics[scale=0.2]{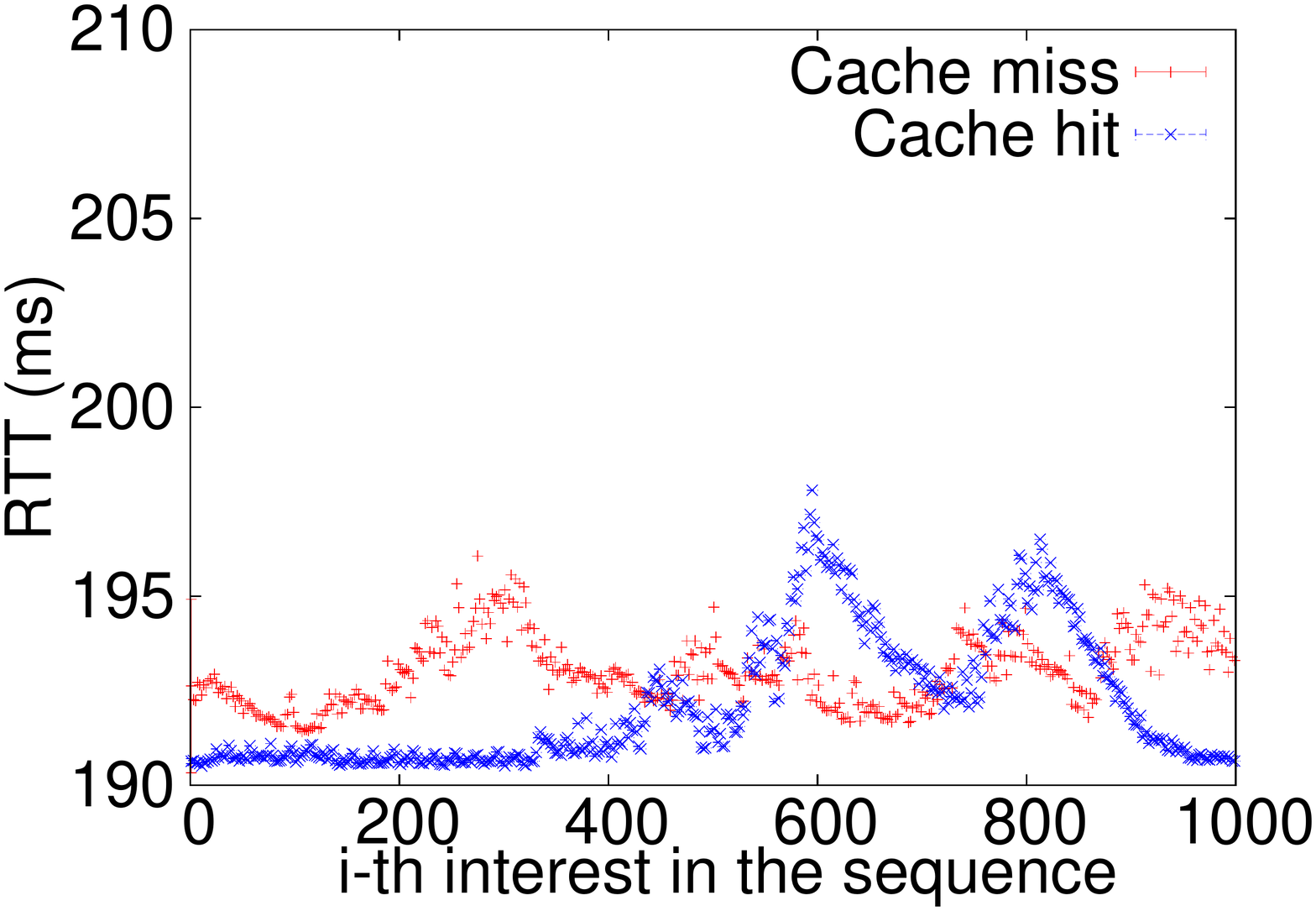}
		            \caption{Testbed, $t=0.5$ ms}
		            \label{img:cache_hit_miss-0.5_95_TB}
        		\end{subfigure}
        		\caption {RTT for data packets, varying request rate.
		\label{img:cache_hit_miss_burst}}
\end{figure*}

We then looked into how interest sending rate affects RTT. We selected values 
for $t$ varying from $t_\mathit{min} = 0.3\ \mu$s to $t=5$ ms (see Section~\ref{sec:multi-bit}). We performed 
several experiments, each using 100,000 data packets. Before each experiment, we restarted 
\router\ in order to remove all cache entries. Results are reported in 
Figure~\ref{img:cache_hit_miss_burst}.

In LAN (figures~\ref{img:cache_hit_miss-0.0_95_LAN}, \ref{img:cache_hit_miss-0.2_95_LAN}, 
and \ref{img:cache_hit_miss-0.5_95_LAN}), RTTs of cache hits and cache misses are clearly 
separated, regardless of $t$. On the testbed (figures~\ref{img:cache_hit_miss-0.0_95_TB}, 
\ref{img:cache_hit_miss-0.2_95_TB}, and \ref{img:cache_hit_miss-0.5_95_TB}), for small values 
of $t$,  cache hits and misses significantly overlap for messages longer than 200 bits. 
This suggests that short busts, separated by short pauses, provide lower error rates.

		\begin{figure}[ht]
			\begin{subfigure}[b]{0.48\textwidth}
		            \includegraphics[scale=0.25]{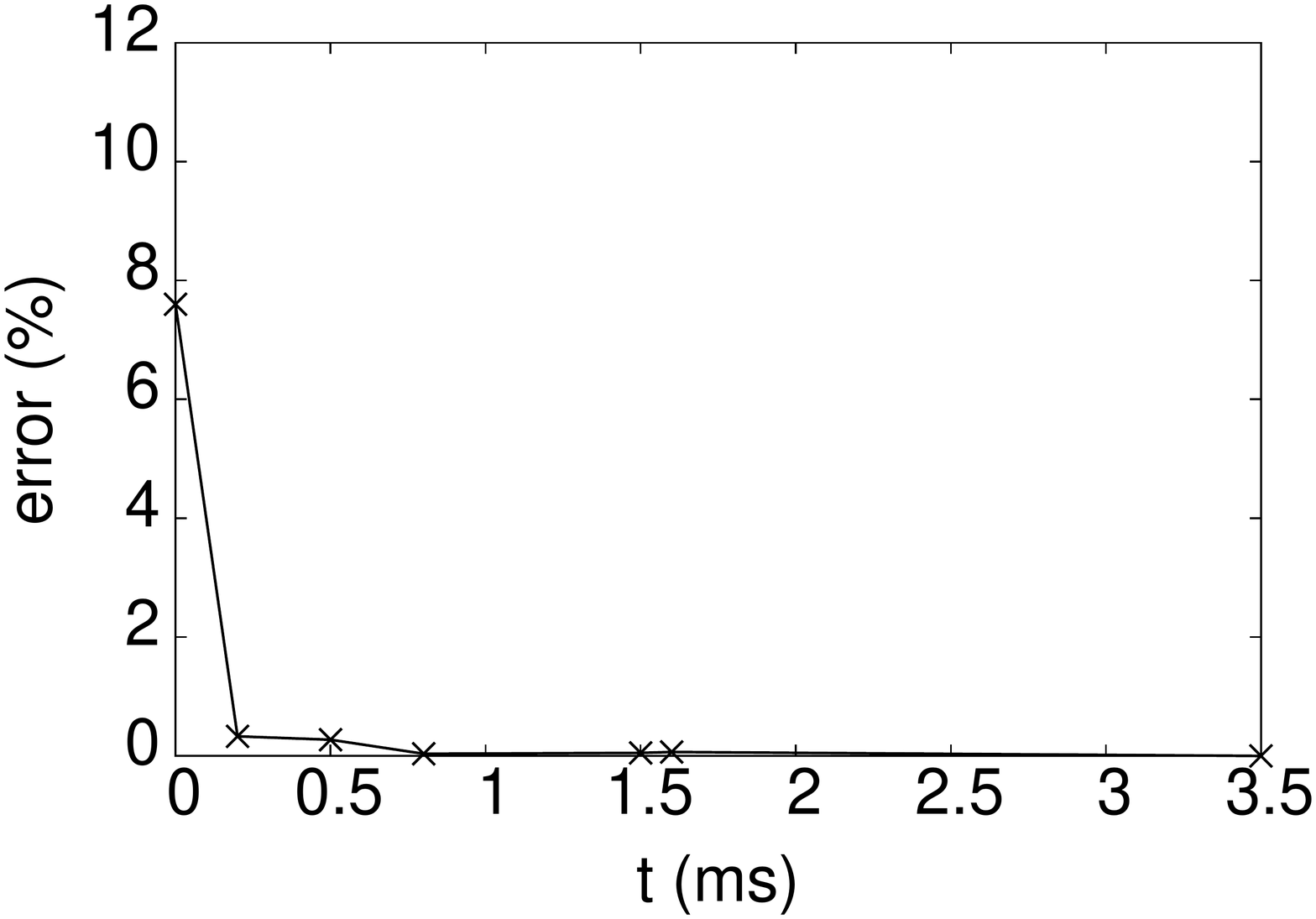}
		            \caption{LAN}
		             \label{img:sct_cache_receiver_t_local}
        		\end{subfigure}
		
			\vspace{-0.4cm}
        	
		\begin{subfigure}[b]{0.48\textwidth}
		            \includegraphics[scale=0.25]{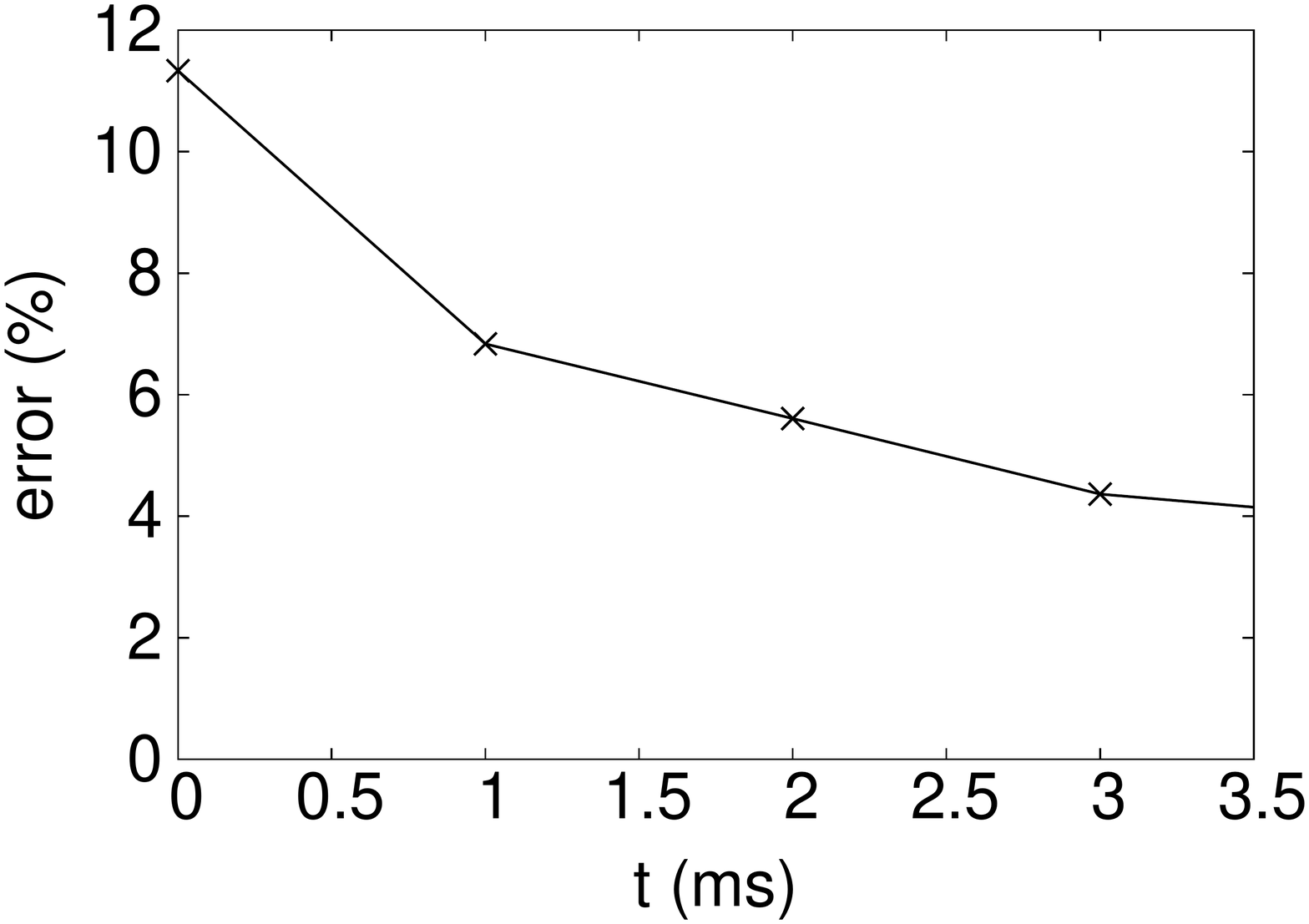}
		            \caption{Testbed}
		             \label{img:sct_cache_receiver_t_testbed}
        		\end{subfigure}
        		\caption {Cache-hit-based communication: write error, varying $t$.\label{img:sct_cache_sender}}
	\end{figure}

	\begin{figure}[ht]
			\begin{subfigure}[b]{0.48\textwidth}
		            \centering
		            \includegraphics[scale=0.5]{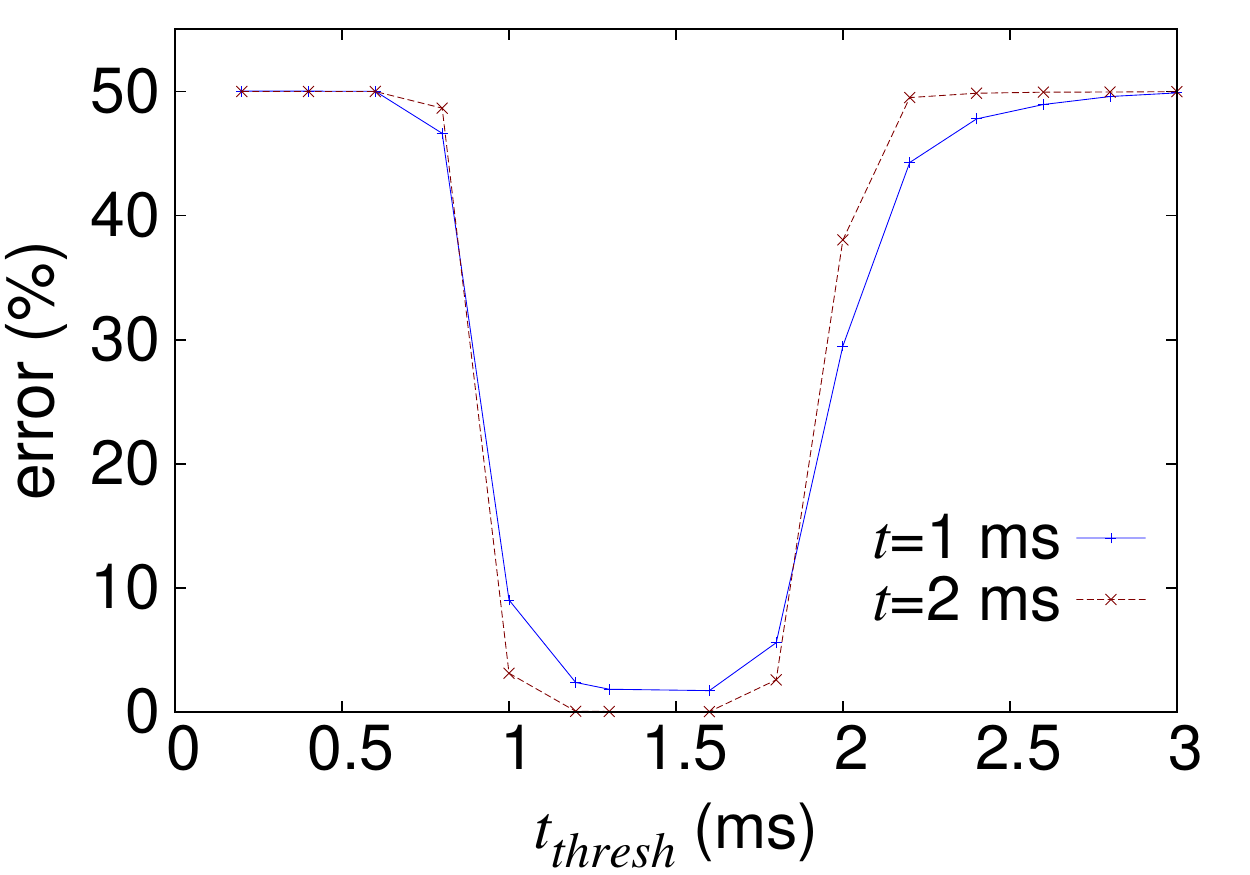}
		            \caption{LAN}
		             \label{img:sct_cache_receiver_local}
        		\end{subfigure}

			\vspace{-0.4cm}

        		\begin{subfigure}[b]{0.48\textwidth}
		            \centering
		            \includegraphics[scale=0.5]{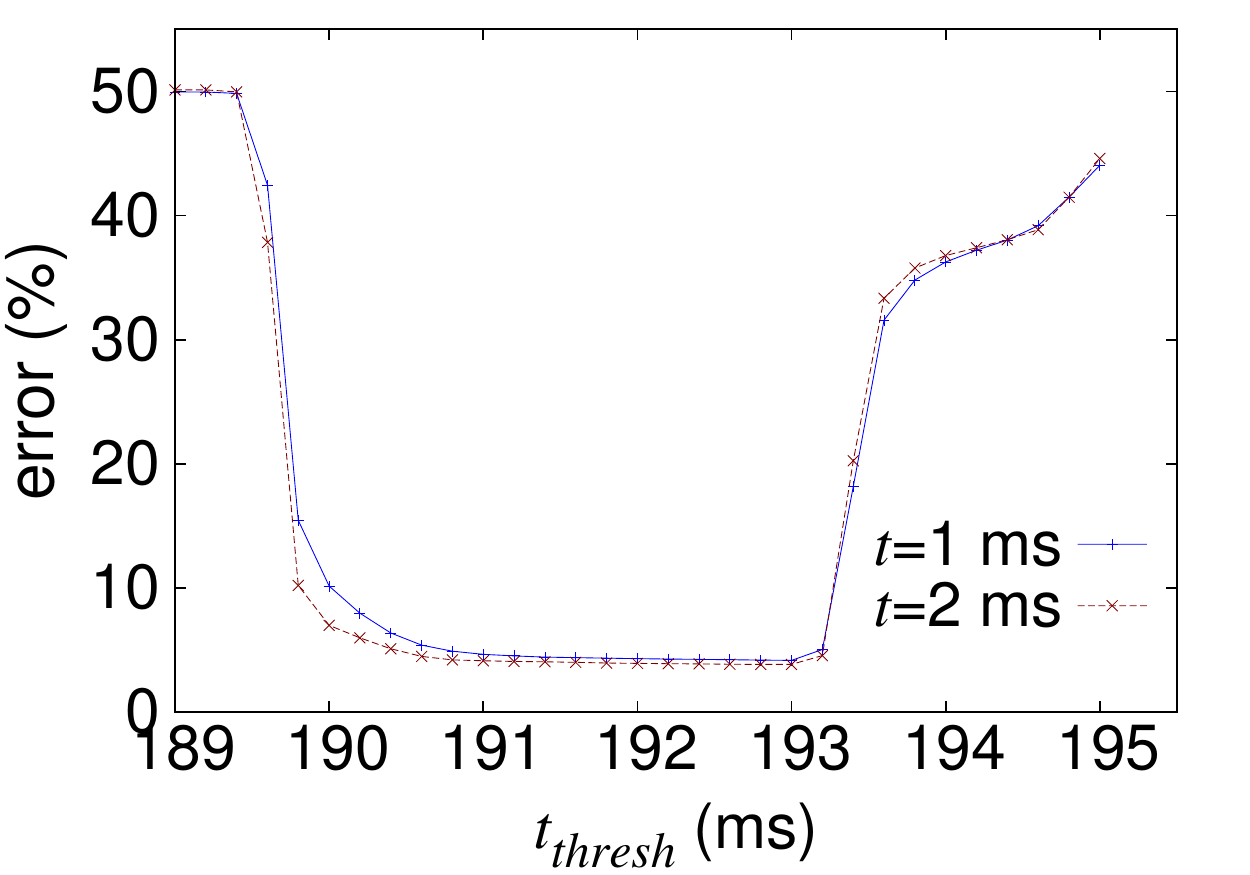}
		            \caption{Testbed}
		             \label{img:sct_cache_receiver_testbed}
        		\end{subfigure}
        		\caption {Cache-hit-based communication: read error varying $t_\mathit{thresh}$ and $t$. \label{img:sct_cache_receiver}}
	\end{figure}

For cache-based CEC, we evaluated read and write errors separately, while varying $t$ and 
$t_\mathit{thresh}$. 
To evaluate write errors, \sender\ published of 100,000 covert bits for each value of $t$. 
Covert bits were subsequently requested at a low rate ($t = 
100$ ms) by \receiver. We then estimated how many data packets were not retrieved from cache. 
Figure~\ref{img:sct_cache_sender} summarizes our findings. 
In this experiment, \receiver\ introduces a small measurement error. We estimate to be 
negligible in LAN, and below 1.5\% on the testbed. With cache-based CEC, write errors can be completely 
eliminated if \sender\ re-issues interests for content that it did not receive; although, 
writing time increases.

To measure read errors, \sender\ published 100,000 covert bits, separated in groups of 1,000-
bit CEMs, for each value of $t$ and $t_\mathit{thresh}$. 
Results of this experiment are shown in Figure~\ref{img:sct_cache_receiver}. 
Due to the clear separation between RTTs associated with cache hits and misses in LAN, read 
errors were very low for a wide range of parameters (e.g., for $t_\mathit{thresh}$ between 1 and 1.5 ms). On the testbed, error was typically between 3\% and 5\% for $t_\mathit{thresh}$ between 191 and 193 ms.

\subsection{Evaluation of Delay-Base (PIT) Techniques} 

\begin{figure}[ht]
			\begin{subfigure}[b]{0.48\textwidth}
		            \includegraphics[scale=0.25]{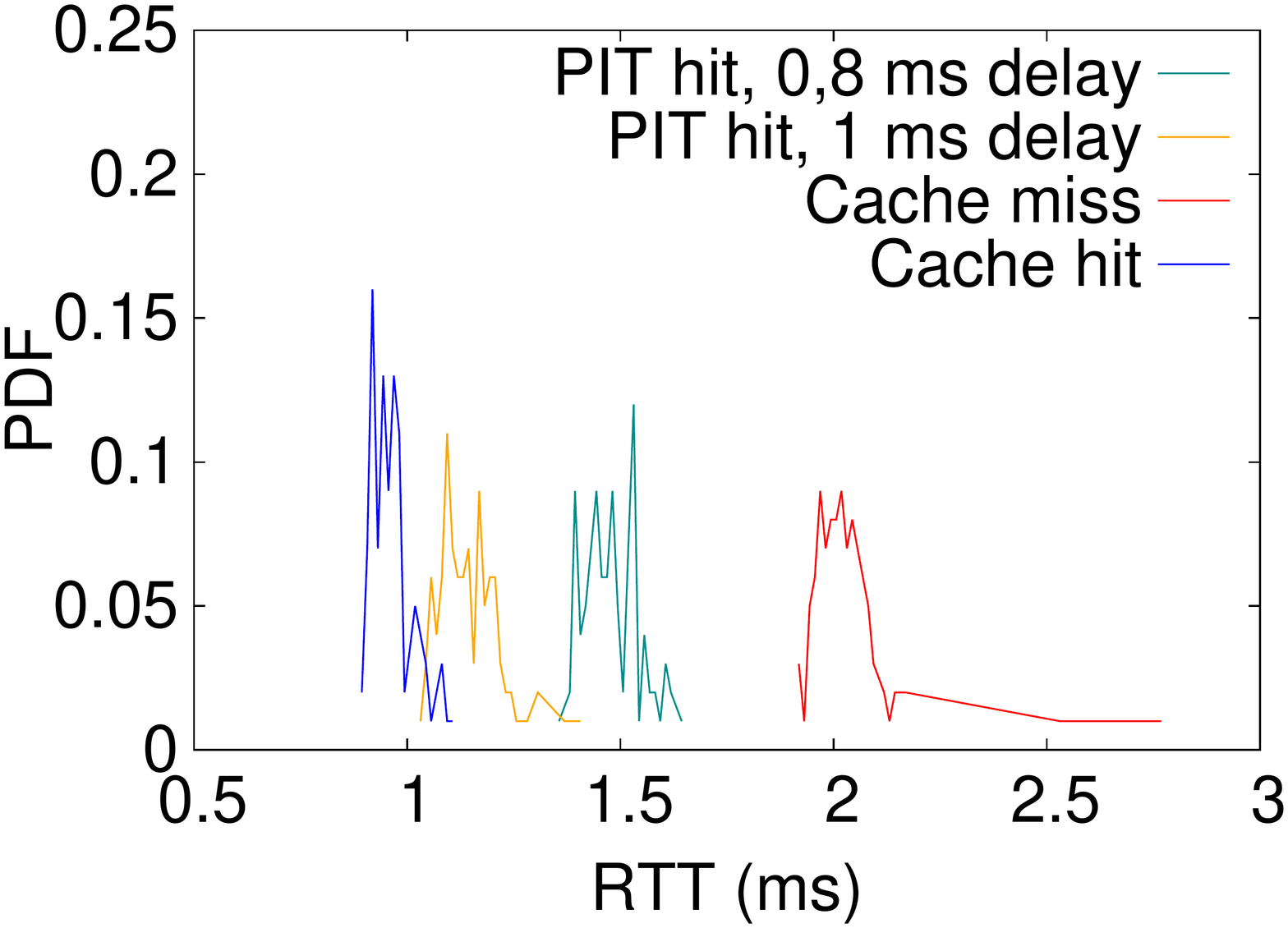}
		            \caption{LAN}
		            \label{img:pit_resp_time_distr_LAN}
        		\end{subfigure}
		
			\vspace{-0.4cm}
        	
		\begin{subfigure}[b]{0.48\textwidth}
		            \includegraphics[scale=0.25]{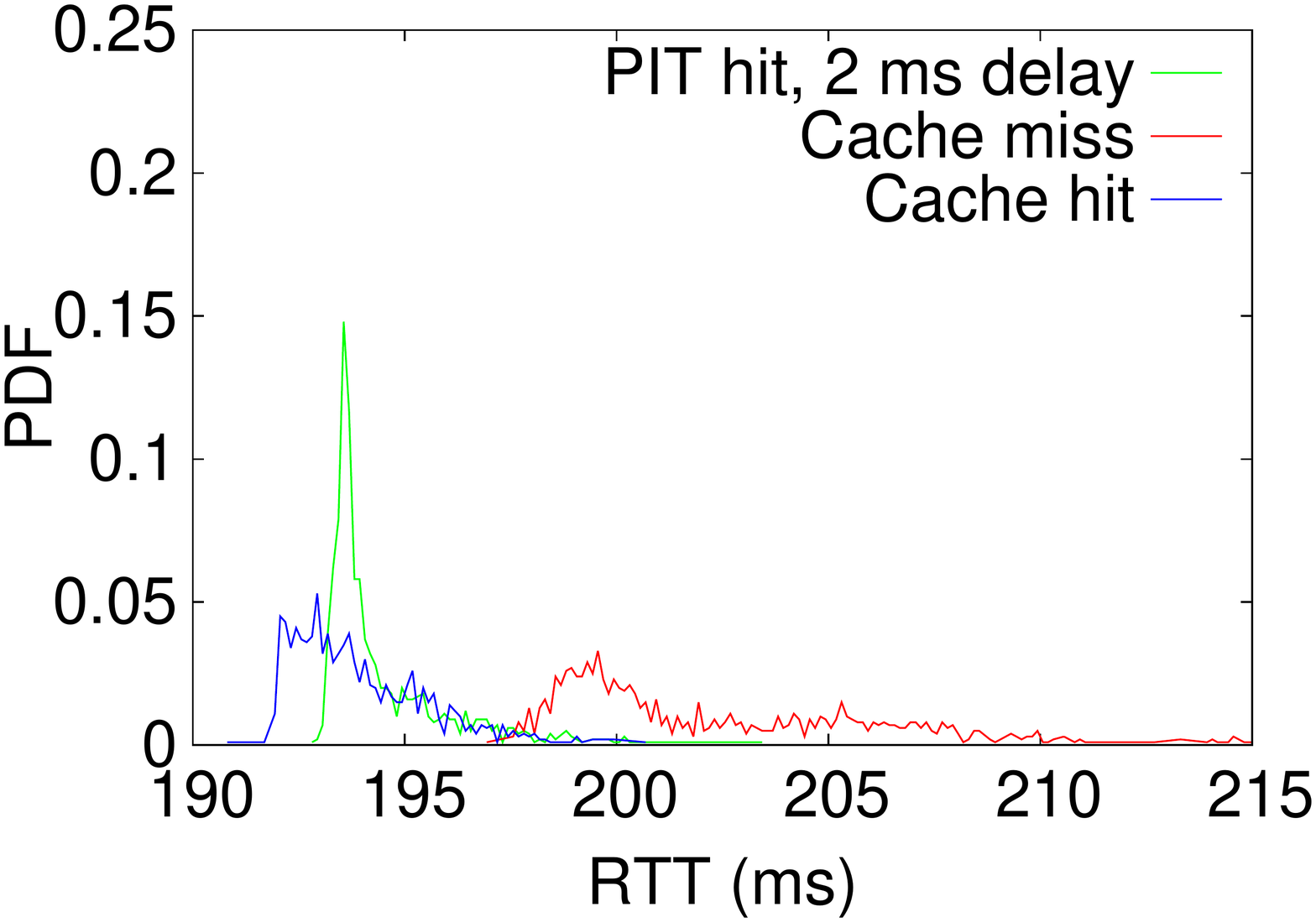}
		            \caption{Testbed}
		            \label{img:pit_resp_time_distr_Testbed}
        		\end{subfigure}
        		\caption {RTT for data packets causing PIT collisions.\label{PIT_resp_time_ALL}}
	\end{figure}

We 
requested the same data packet from both \snr\ at very close intervals (i.e., 0.8 and 1~ms in 
 LAN and 2 ms on testbed), in order to trigger interest collapsing on \router, and, 
therefore, a PIT hit. \snr\ were synchronized using a local NTP server; we 
estimated the time difference between the two hosts to be below 0.2 ms. Our experiments show that is 
possible to distinguish PIT hits from misses using appropriate intervals between interests 
from \snr. Results of this 
experiment are shown in Figure~\ref{PIT_resp_time_ALL}. 
However, the separation is less clear than with cache, as shown in the same figure. 
Moreover, this channel requires much tighter synchronization between \snr\ (i.e., 
sub-millisecond in LAN, and within 2 ms on testbed). For these reasons, 
PIT-based CEC are significantly more difficult to implement.

\begin{figure}[ht]
			\begin{subfigure}[b]{0.48\textwidth}
		            \centering
		            \includegraphics[scale=0.5]{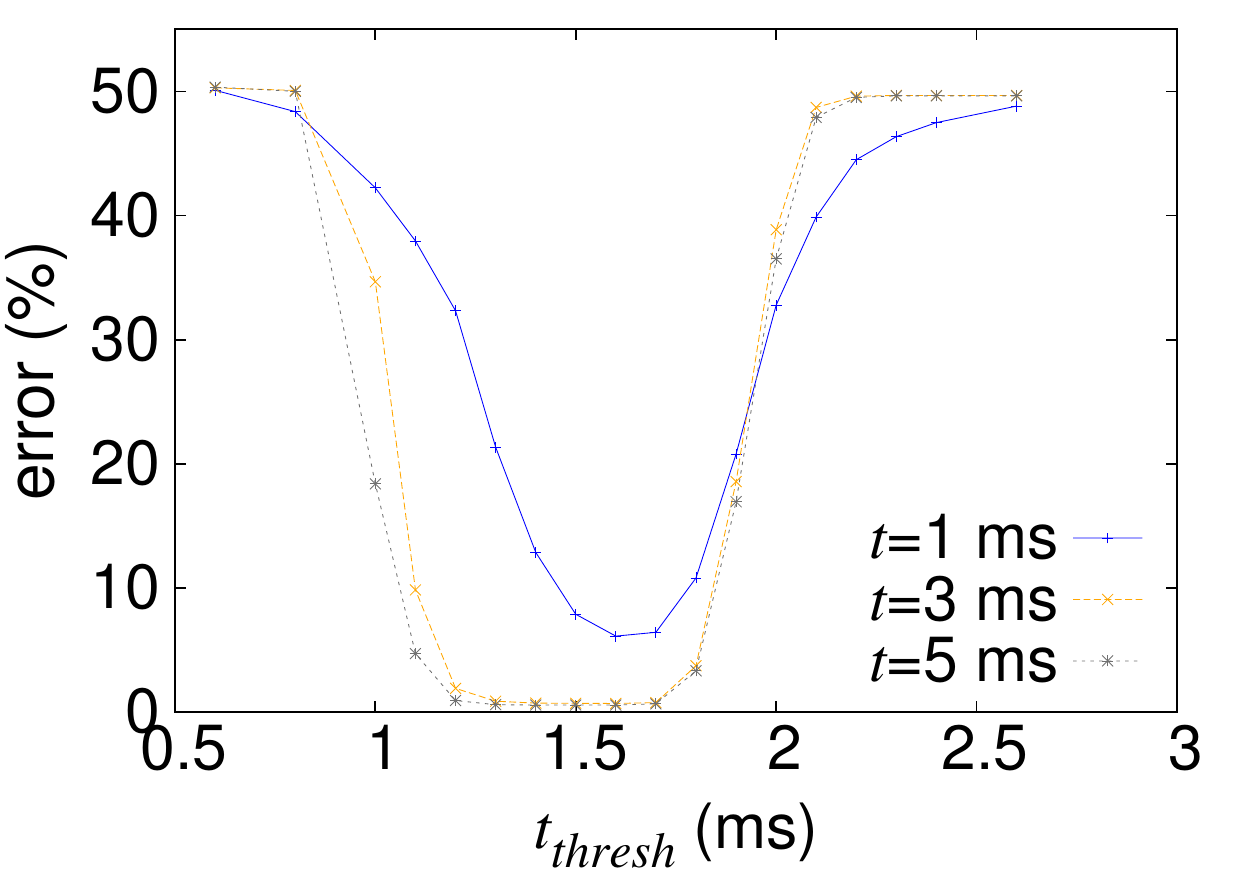}
		            \caption{LAN}
		            \label{img:sct_pit_local}
        		\end{subfigure}
			
        		\begin{subfigure}[b]{0.48\textwidth}
		            \centering
		            \includegraphics[scale=0.5]{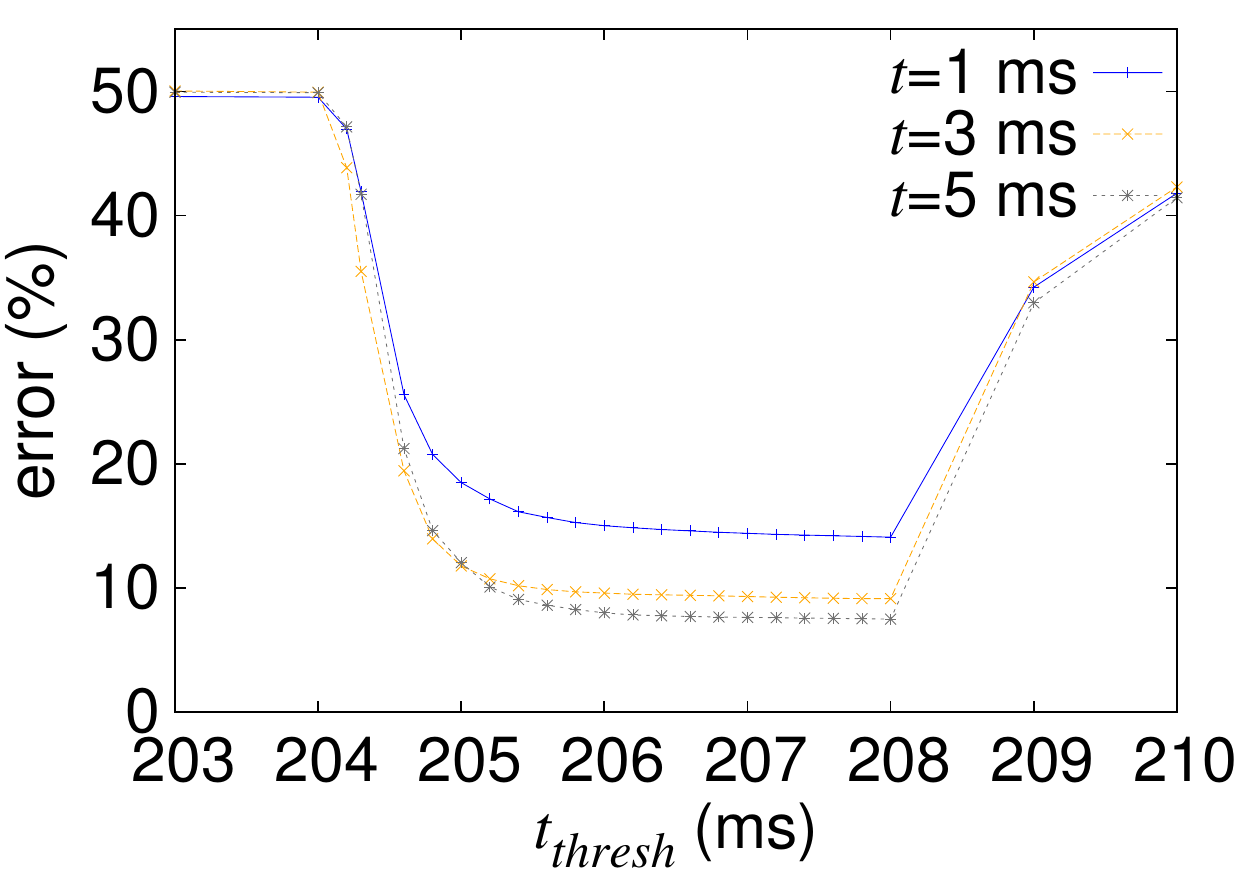}
		            \caption{Testbed}
		            \label{img:sct_pit_testbed}
        		\end{subfigure}
        		\caption {Joint write and read error varying $t$ in our PIT hit-based protocol. \label{img:sct_pit}}
	\end{figure}

Since \snr\ must operate synchronously and with the same $t$, we measured read and write errors  
jointly. For this experiment, the delay between interests from \snr\ is 0.8 ms in LAN, and 8 ms 
on the testbed. Results are shown in Figure~\ref{img:sct_pit}. 
With appropriate 
choice of the threshold parameter, errors in LAN are negligible, and below 7.5\% in the 
testbed.

\subsection{TDP Evaluation}

\begin{figure}[ht]
				\begin{subfigure}[b]{0.4\textwidth}
		            \includegraphics[scale=0.25]{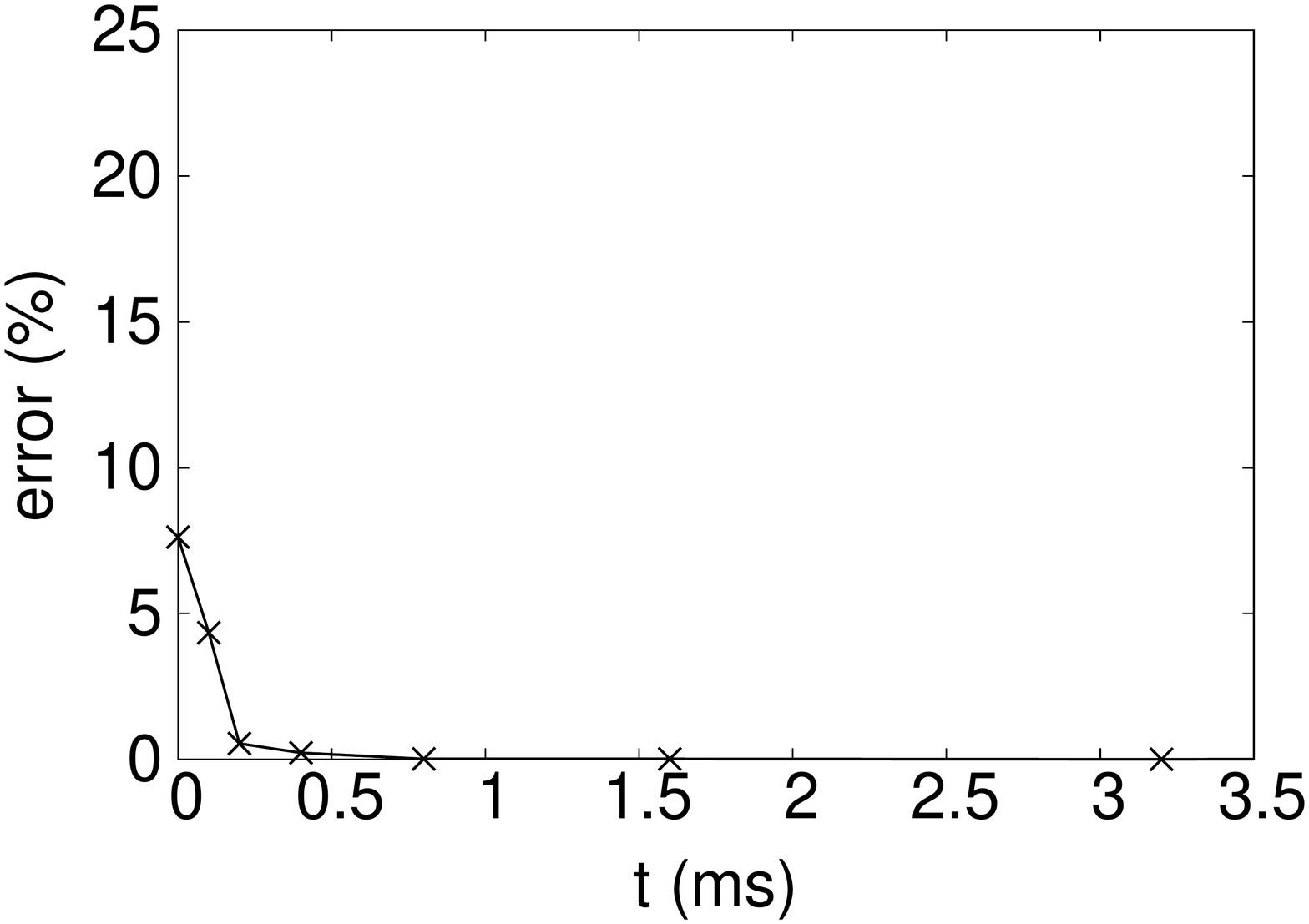}
			    	\label{img:nmsp-sender-local}
		            \caption{LAN}
        		\end{subfigure}
				\hspace{0.4cm}
        		\begin{subfigure}[b]{0.4\textwidth}
		            \includegraphics[scale=0.25]{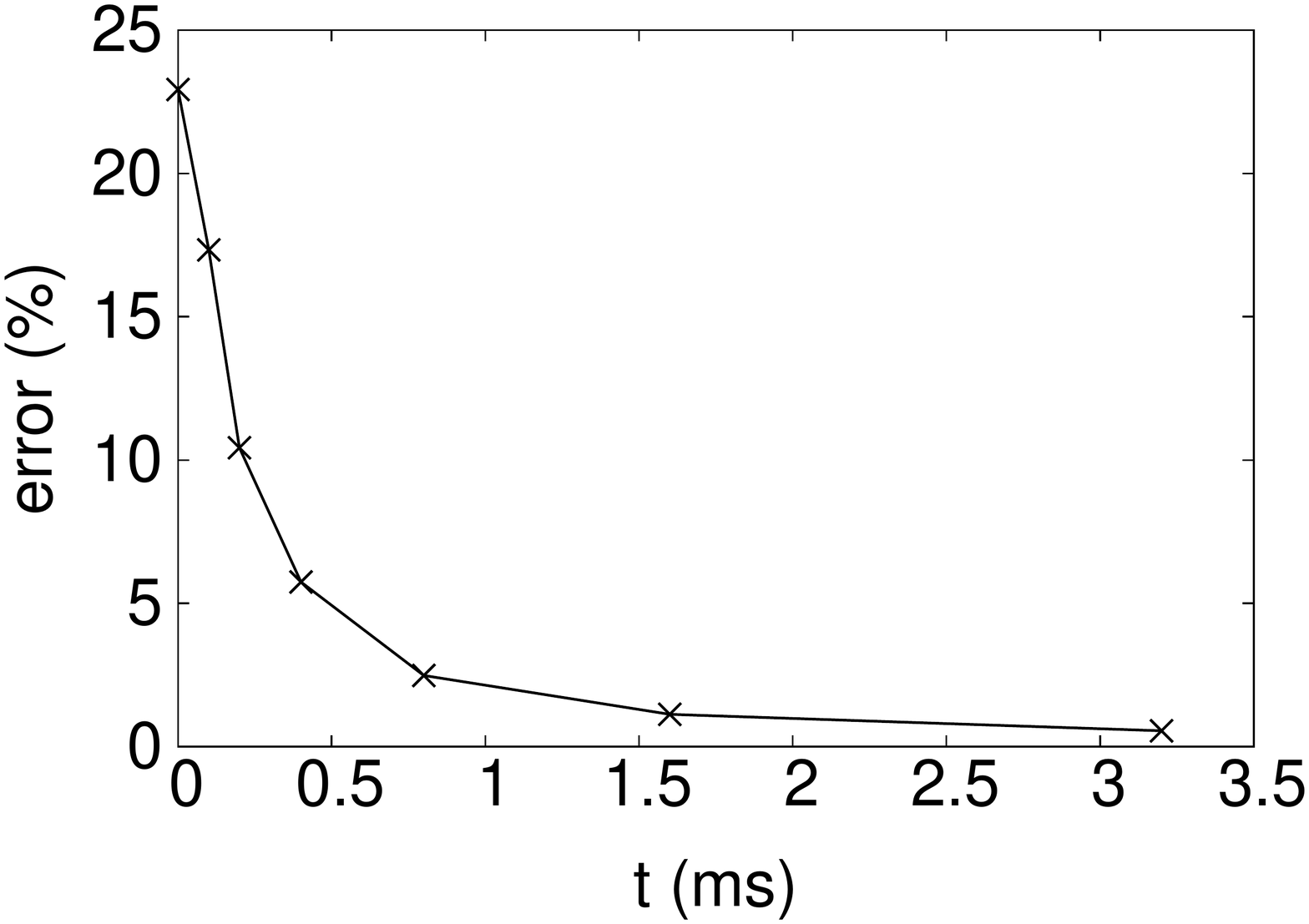}
			    	\label{img:nmsp-sender-testbed}
		            \caption{Testbed}
        		\end{subfigure}        		
        	\caption{Write error with TDP, varying \sender's $t$.\label{img:nmsp-receiver-TDP}}
	\end{figure}

\begin{figure}[h!]
			\begin{subfigure}[b]{0.4\textwidth}
		            \includegraphics[scale=0.25]{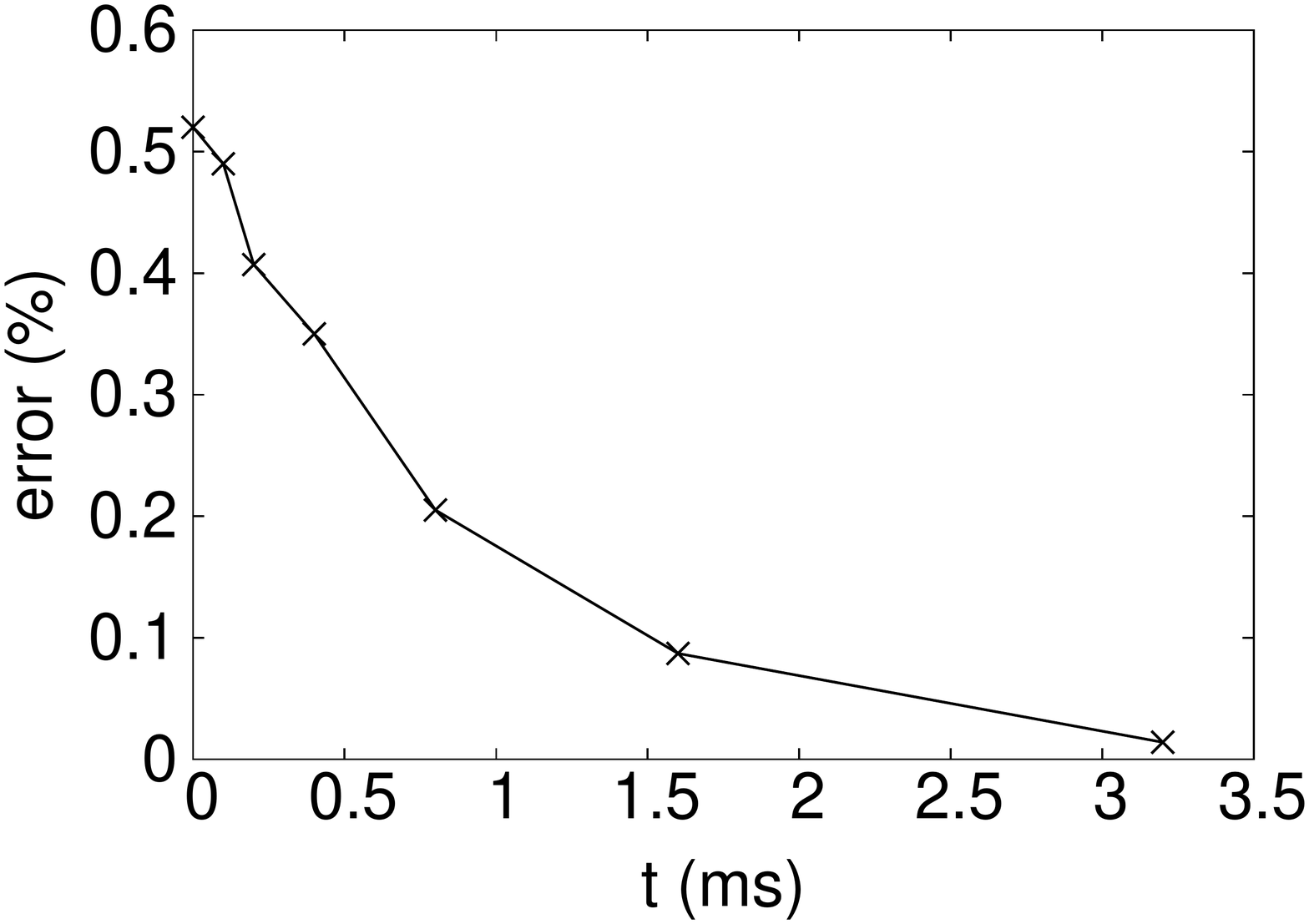}
		            \caption{LAN}
		            \label{img:res_sct_burst_cache_LAN}
        		\end{subfigure}
			\hspace{0.4cm}
        		\begin{subfigure}[b]{0.4\textwidth}
		            \includegraphics[scale=0.25]{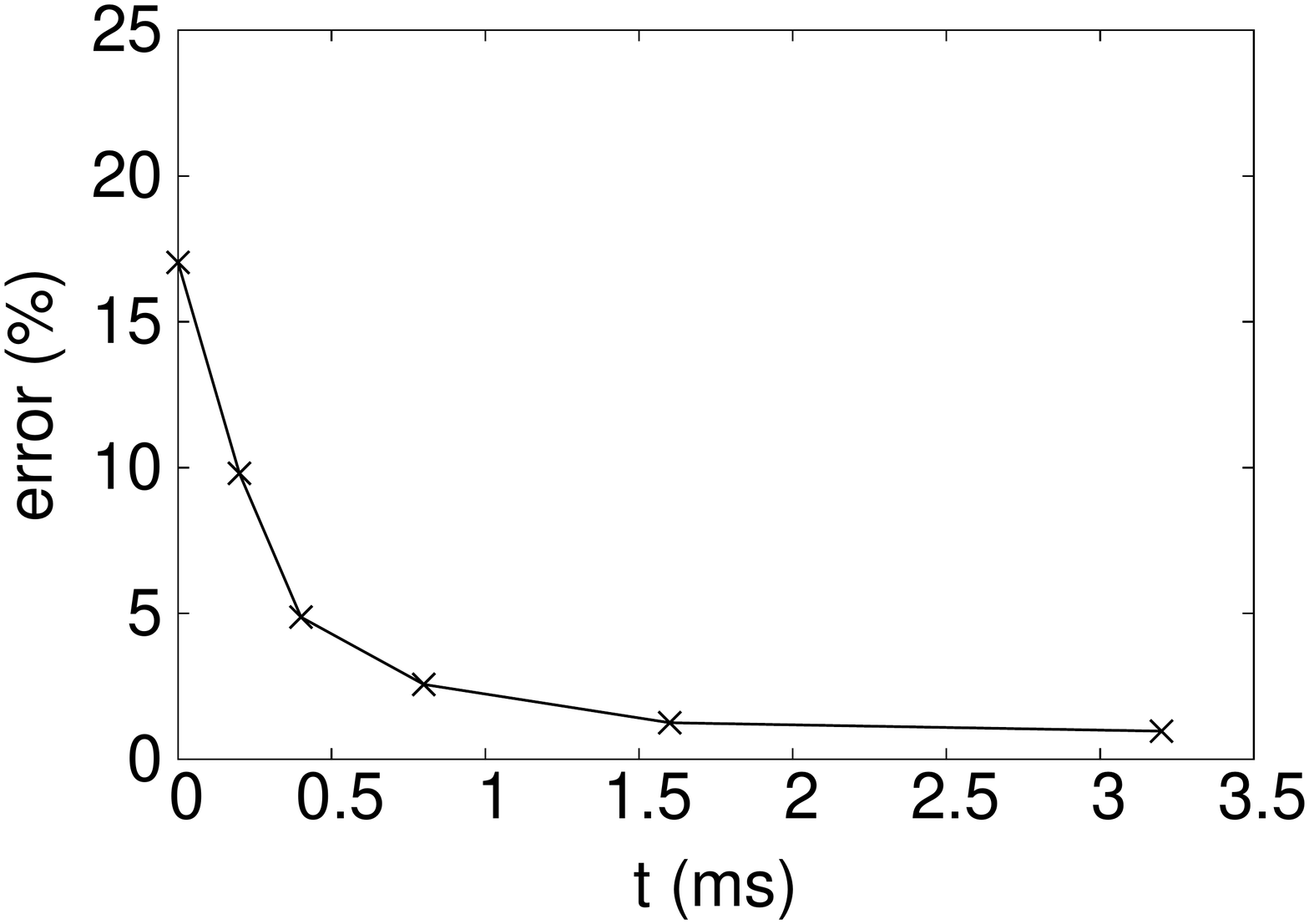}
		            \caption{Testbed}
		            \label{img:res_sct_burst_cache_testbed}
        		\end{subfigure}
        		\caption {Read error with TDP, varying \receiver's $t$.\label{img:res_sct_burst_cache}}
	\end{figure}

We measured the error rate varying write and read speeds separately for \snr. Figures~\ref{img:nmsp-receiver-TDP} and \ref{img:res_sct_burst_cache} summarize our findings. On the receiver side, this technique performs better than the cache-hit-based one. For example, for $t$ = 1.5 ms in the testbed, the error for TPD is less than 2\% (see Figure \ref{img:res_sct_burst_cache_testbed}), while for $t=3$ (i.e., the same effective bit rate relative to the CEM) in the cache-hit-based technique  the error for is more than 4\% (Figure \ref {img:sct_cache_receiver_t_testbed}).

\subsection{Evaluation of Common-Prefix-Based Technique} 

We set $m=1$ (i.e., each data packet encodes one 
bit), in order to encode 1,000-bit CEM using 1,000 data packets.
We run separate experiments to evaluate \snr\ errors. As mentioned in 
Section~\ref{sec:error_handling}, both parties can avoid packet-loss-induced errors 
using interest retransmission. For a fair comparison with previous protocols, we test how the 
common-prefix-based technique performs {\em without} retransmissions.

\begin{figure}[ht]
	\centering
    \includegraphics[scale=0.25]{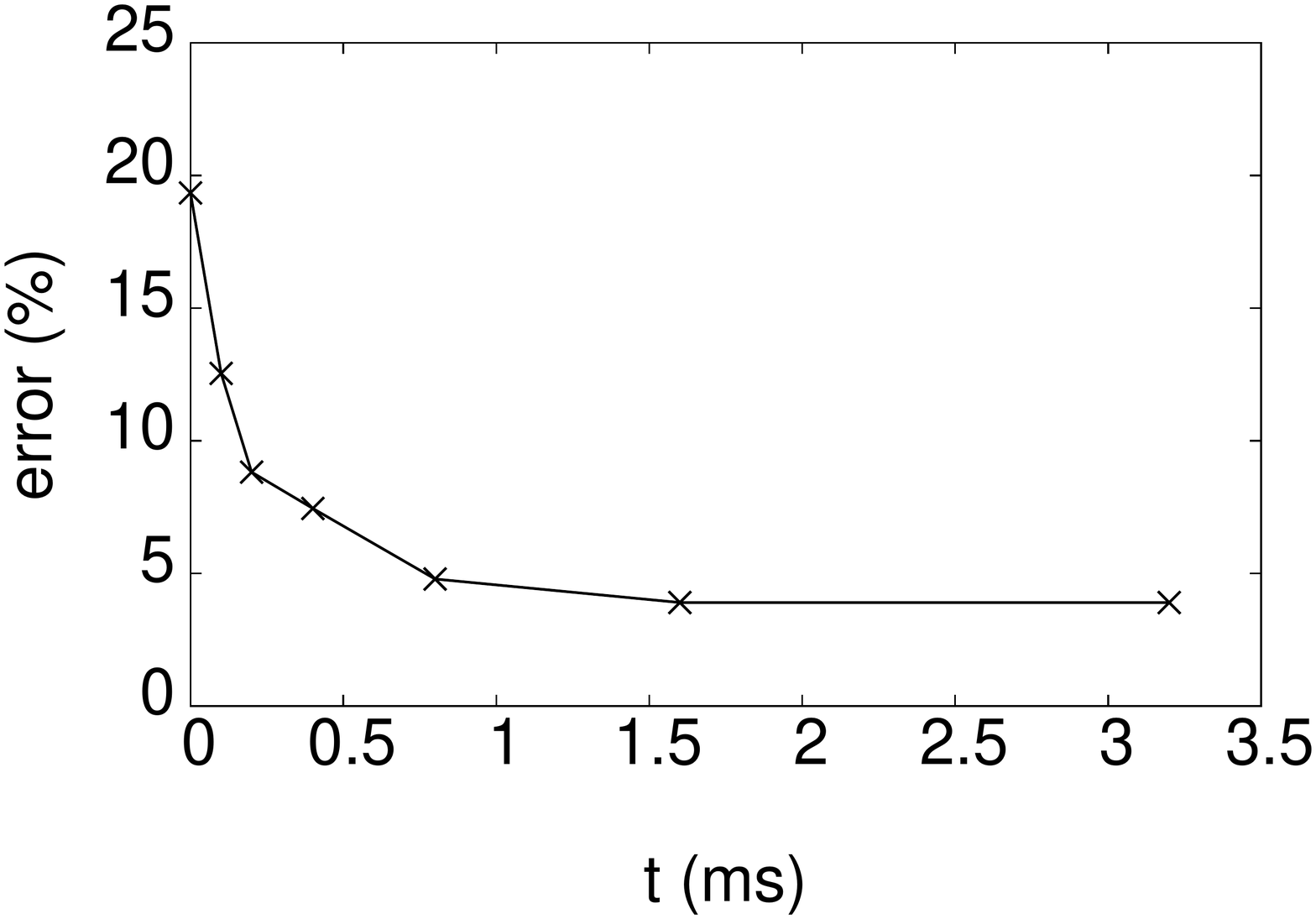}
    \caption{Common-prefix-based protocol: read error varying $t$.\label{img:nmsp-receiver}} 
\end{figure}

Results on write errors, both in our LAN and on the testbed, are identical to those in 
Figure~\ref{img:res_sct_burst_cache}. In fact, \sender\ performs the 
same actions to send a CEM.
Read errors on the testbed are reported in Figure~\ref{img:nmsp-receiver}. We omit the plot corresponding to read errors in LAN, since for all tested values of $t$ error rate was below 0.03\%. Errors for both \snr\ are due to packet loss.

\subsection{Bit Rate and Error Comparison}
\begin{figure*}[ht!]
			   	\centering
			   	\begin{subfigure}[b]{0.48\textwidth}
					\includegraphics[scale=0.3]{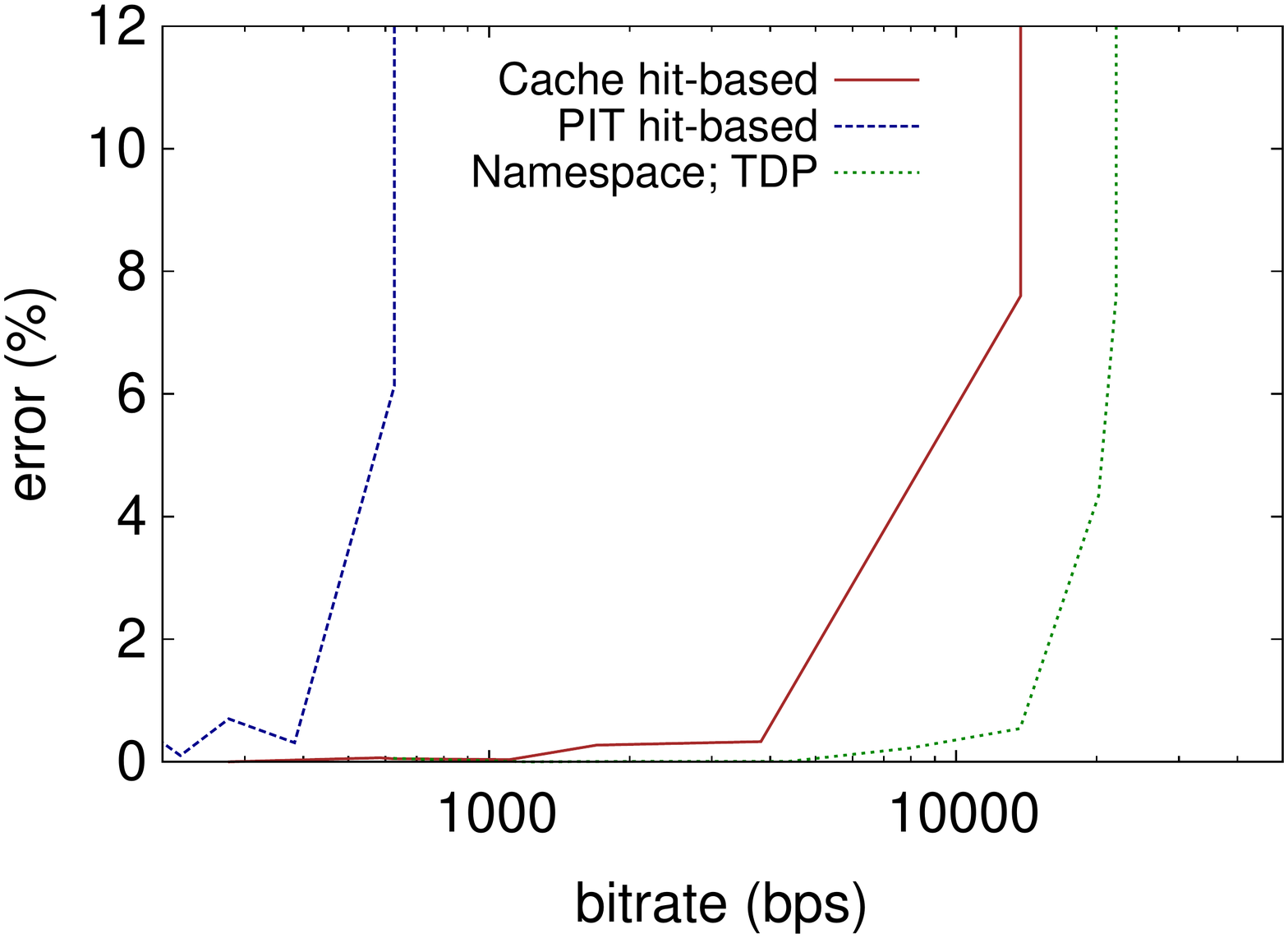}
					\caption{LAN (\sender)} 
				\end{subfigure}
			   	\begin{subfigure}[b]{0.48\textwidth}
					\includegraphics[scale=0.3]{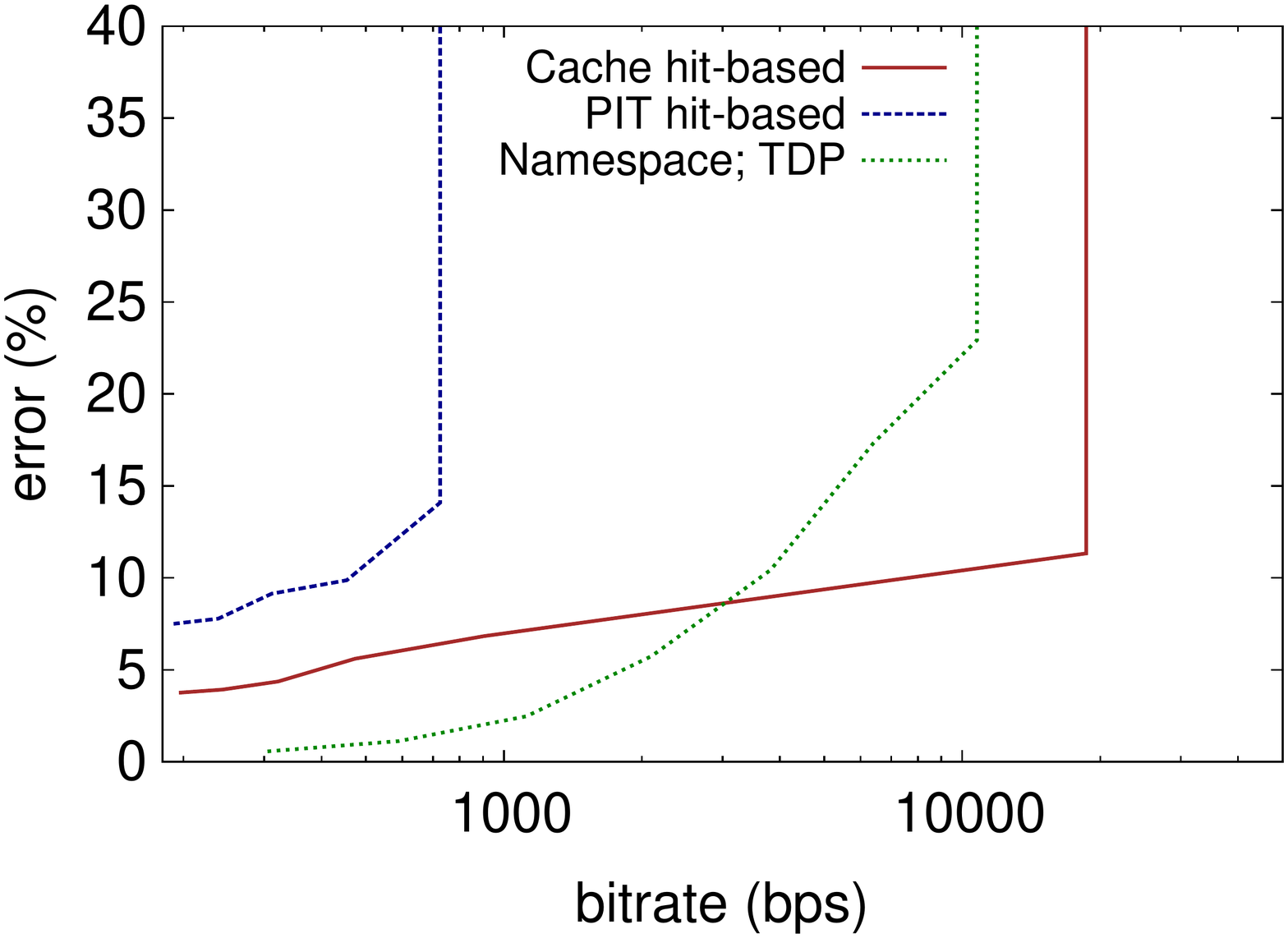}
					\caption{Testbed (\sender)} 
				\end{subfigure}
				 	\\
				\begin{subfigure}[b]{0.48\textwidth}
					\includegraphics[scale=0.3]{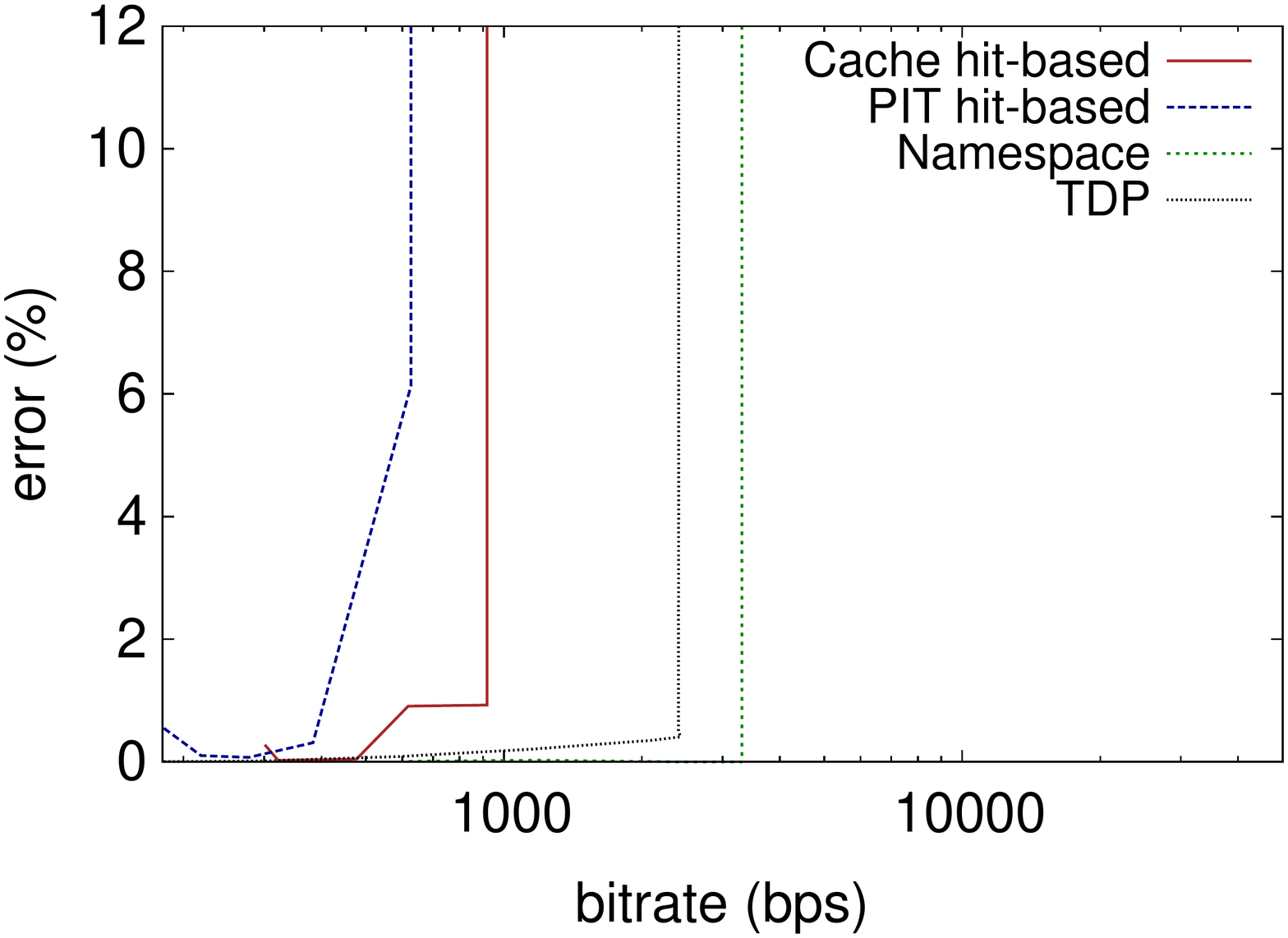}
					\caption{LAN (\receiver)} 
				\end{subfigure}
				\begin{subfigure}[b]{0.48\textwidth}
					\includegraphics[scale=0.3]{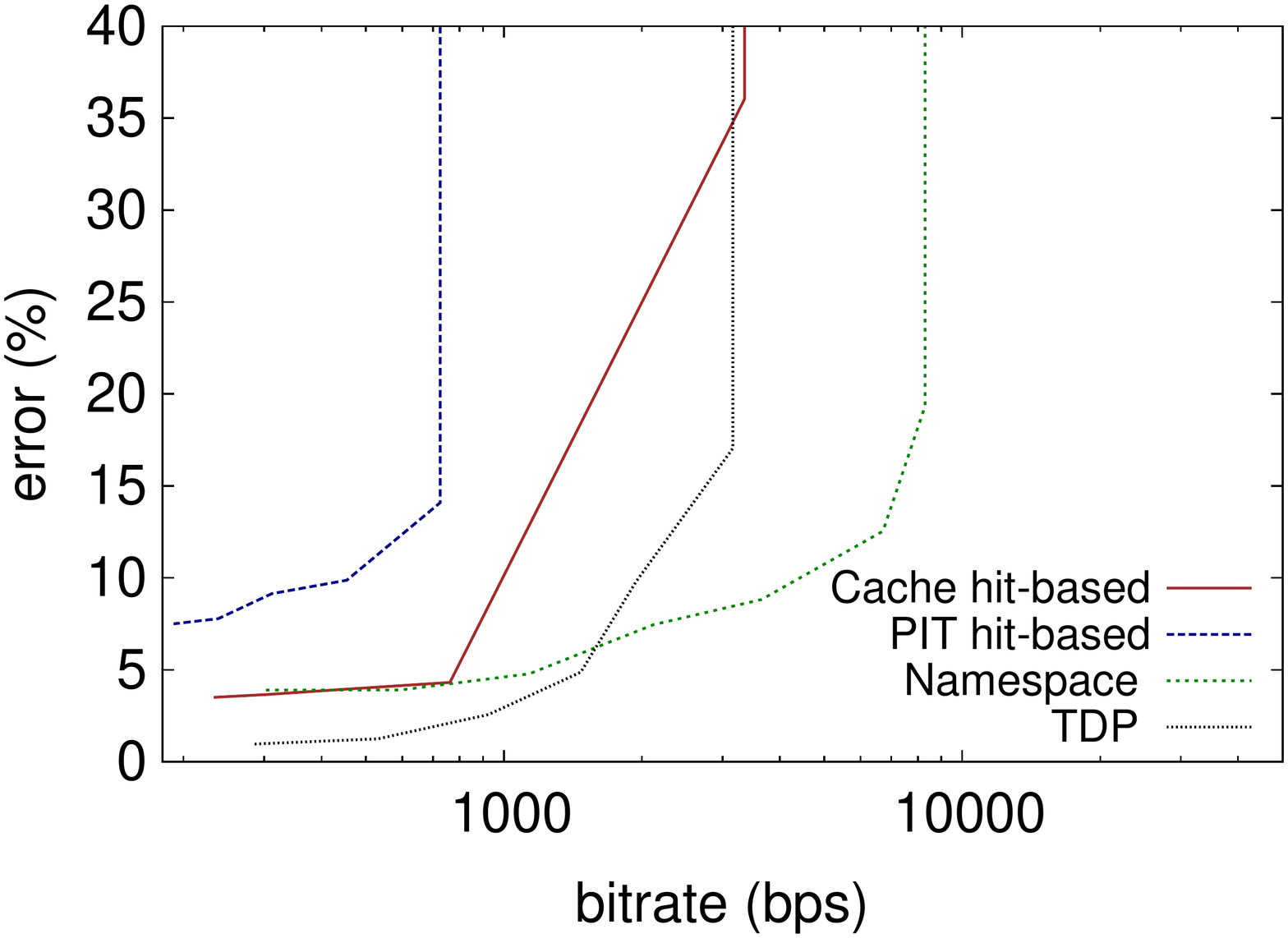}
					\caption{Testbed (\receiver)} 
				\end{subfigure}
				\caption{Performance comparison.}
			\label{img:FinalComparison}
				
			\end{figure*}

To simplify comparison of techniques introduced in this paper, we combine effective bit 
rate and corresponding error for all our protocols in Figure~\ref{img:FinalComparison}. Note 
that, for TDP, \sender's effective bit rate can be {\em multiplied} by an arbitrary $m$, while 
\receiver's bit rate should be {\em divided} by $2^m$. Analogously, the bit rate for both \snr\ in the common-prefix protocol should be multiplied by $m$ as discussed in Section~\ref{sec:namespacechannel}.

\section{Security Analysis}
\label{sec:securityanalysis}

We now analyze security of CEC techniques. We start by showing that proposed 
protocols are retroactively private and secure against message recovery attack. We then conclude with an informal  discussion on the detectability and robustness of our approaches. 

\subsection{Retroactive Privacy}

\adversary\ has non-negligible advantage over 1/2 in the retroactive privacy game (see Section~\ref{systemmodel}) only if it can  
infer information about $a$ from interaction with \sender, \receiver\ and 
\router\ {\em after} the message $M_a$ has expired. That is, \adversary\ can only interact with 
protocol participants after data packets used to encode $M_a$ have been removed from 
\router's PIT and from all caches.

Since \snr\ delete $M_a$ as soon as they  (respectively) send and receive it, \adversary\ 
cannot acquire information about $M_a$ by compromising the two parties.
Similarly, NDN routers do not keep track of data packets once they disappear from both 
PIT and cache. Therefore, after $M_a$ expires, \router\ carries no information about the 
message.
As a result, there is simply no information about $M_a$ within the network after the message 
expires. 

\subsection{Security Against Message-Recovery Attacks}

In order to reconstruct a CEM, \adversary\ can probe all NDN routers, and try to identify 
data packets used for covert communication. However, this approach has two problems: (1) 
there is no data packet in routers caches for a bit set to 0; therefore, \adversary\ cannot learn 
information about these bits by simply observing routers caches. (2) even for a relatively small NDN deployment, the number of routers and the size of their caches makes this attack infeasible.

Another adversarial strategy consists in infiltration of the routing infrastructure: 
\adversary\ 
could mount a Sybil attack~\cite{defeating-vanish}, deploying a large number of malicious NDN routers. We believe that 
this approach is not feasible, since: (1) \adversary\ cannot  deploy an 
arbitrary number of NDN routers. Even if NDN is implemented as an overlay, routers are identified by 
their unique IP address. This would force \adversary\ to obtain a very large number of public 
IP address. (2) Even if the adversary succeeds deploying a large number of routers, it must log all data packets forwarded by all controlled routers. This may not be feasible. (3) Similarly, even if \adversary\ can compromise arbitrary routers, maintaining logs for all forwarded data packets would not be viable.

\subsection{Detectability}

In order to exchange a message through our protocols, \snr\ do not need to communicate directly, 
nor they need to be connected through the same NDN router. Moreover, they only interact with the 
network as prescribed by NDN specifications.

A single-bit message $b=0$ sent using single-bit transmission via cache or PIT cannot be 
detected, since \sender\ performs no action. When $b=1$, \sender\ retrieves a non-popular 
data packet. We believe that, in practice, by flagging all single interests for non-popular 
data packets as ``suspicious'', \adversary\ would incur in an overwhelmingly large number of false
alarms.
Similarly, a single interest issued by \receiver\ to retrieve $b$ would be easily hidden by the 
existing traffic.

When \snr\ exchange messages longer than a single bit, however, their actions become more detectable. In particular, 
the longer the message, the more likely it is for \adversary\ to correctly identify a 
CEM between two or more parties. While a single interest for non-popular data packets may not raise any suspect, a long streak of interests for non-popular data packets 
may be easy to notice. For this reason, \snr\ should limit the size of the exchanged messages 
to reduce detectability. 

Finally, with namespace-based covert communication detectability mostly depends on $m$ and on the 
size of the covert. In particular, a higher value for $m$ implies lower detectability: less data packets have to be requested to write and read a covert message.

\subsection{Robustness}
When \router\ introduces arbitrary delays to conceal cache hits, our techniques based on 
measuring time difference between these two events do not work. However, techniques based on PIT 
and on common prefixes are not affected by cache hit delays, since they either do not rely on 
cache or do not consider RTT. 

Similarly, when the network introduces unpredictable delays on packets (e.g., when traffic 
intensity has sudden wide fluctuations), common-prefix-based technique may be more appropriate 
since it does not rely on timing measurements.

\section{Related Work}
\label{sec:related_work}

We divide relevant related work in two classes: {\em covert communication} and {\em ephemeral communication}.

\paragraph{Covert Communication}
The goal of a covert channel is to conceal the very existence of a covert message by 
communicating it through legitimate channels~\cite{CoCo}.

In~\cite{jitterbug}, Shah et al.~present Jitterbug, a hardware device and a communication 
protocol that covertly transmit data by perturbing the timing of keyboard events. In 
particular, the authors design and implement a small hardware {\em pass-through} device that 
introduces small -- although, measurable -- variations in the times at which keyboard events 
are delivered to the host. When the user runs an interactive communication protocol (e.g., SSH, 
instant messaging), a receiver monitoring the host's network traffic can recover the leaked 
data. According to the experimental results reported in~\cite{jitterbug}, the bandwidth offered 
by Jitterbug is roughly 500 bps over 14 network hops, with 8.9\% error rate. In contrast, our 
technique provide a bit rate of about 15,000~bps in a similar scenario with analogous error 
rate. Another difference is that with Jitterbug the receiver must be able to intercept network 
traffic, while our approach can be used by any unprivileged user.

CoCo, introduced in~\cite{CoCo} by Houmansadr et al., is a framework for establishing covert 
channels via inter-packet delays. The sender generates a traffic flow directed to the receiver, 
then manipulates the flow according to the covert message and a key, shared between the two 
parties. The coding algorithm used in CoCo ensures robustness of the covert message to 
perturbations. The authors show statistical evidence on the undetectability of the 
communication channel. We emphasize that CoCo would not satisfy our requirements because sender 
and receiver must communicate directly.

Murdoch et al.~\cite{MurdochL} investigate covert channel implemented by embedding information 
in random-looking TCP fields. They show that na\"ive approaches -- such as embedding ciphertext 
in the initial sequence number (ISN) field -- can be easily detected. Then, they discuss how to 
implement networking stack-specific covert channel, which are provably undetectable. Similarly to CoCo, the main difference between our work and the work of Murdoch et al.~is that sender and receiver must exchange packets directly.

\paragraph{Ephemeral Communication}
Geambasu et al.~introduced the Vanish system~\cite{vanish}, which allows users to publish 
ephemeral messages. Users encrypt their messages using a random symmetric key. Then, they 
publish shares of the key (computed using Shamir secret sharing~\cite{shamirSS}) in random 
indices in a large, pre-existing distributed hash table (DHT). A DHT is a distributed data 
structure that holds key-value pairs. Since data on DHTs is automatically deleted over time, 
shares of the key automatically ``disappear''. Once enough shares have been deleted, the key -- 
and therefore the encrypted message -- is effectively erased.

Wolchok et al.~\cite{defeating-vanish} showed that Vanish can be defeated using low-cost Sybil 
attacks on the DHT. In particular, they exploited one of the design flaws of Vanish, namely the 
assumption that DHTs are resistant to crawling. This is in contrast with our approach, where 
monitoring all routers' caches is clearly infeasible. Although the authors of Vanish have since 
proposed countermeasures~\cite{vanish2}, these techniques 
only slightly raise the bar against existing attacks~\cite{CastellucciaCFK11}.

Castelluccia et al.~\cite{CastellucciaCFK11} introduced EphPub, a DNS-based ephemeral 
communication technique. A publishers encrypts and distributes a message. Then, it distributes 
the decryption key as follows: for each key bit set to 1, the publisher picks a DNS resolver 
and uses it to answer a recursive DNS queries for a specific domain. Since DNS resolvers cache 
responses for a pre-determined amount of time, one or more receivers can subsequently issue 
{\em non-recoursive} queries to the same resolver. These queries will be answered only if the 
corresponding domain-IP pair is in cache. Once enough cache entries expire (or get 
overwritten), the decryption key -- and therefore the published message -- disappears. 

There are several differences between EphPub and our techniques. First, while 
EphPub relies on an application-layer service (DNS resolver) to publish an 
ephemeral piece of data, our techniques leverage routers' PITs and caches, which 
are part of the routing architecture. Moreover, while EphPub can be blocked by 
forcing users to use a local DNS server with no cache (e.g., by filtering out DNS queries at the network gateway), our PIT-based 
technique allows two parties to exchange CEMs even if routers do not provide 
content caching. Moreover, if EphPub sees wide adoption, there are several 
concerns (raised also by Castelluccia et al.~in~\cite{CastellucciaCFK11}) that 
would impose excessive load on DNS servers, which would then be forced to 
stop acting as ``open'' resolvers. In contrast, with our approach, communicating 
parties do not impose higher-than-usual load on routers: consumers simply use 
their allocated bandwidth for content retrieval. 
Furthermore, routers cannot determine the source of data requests (interests do 
not carry a source address), and therefore always operate similarly to open 
resolvers. Finally, EphPub does not provides covert communication, 
since the behavior of two users who communicate via EphPub is difficult to 
conceal. In fact, ``regular'' users rarely query multiple remote DNS servers in 
short bursts. With our techniques, instead, \snr~do not perform any easily 
identifiable activity. 

Perlman~\cite{ephemerizer} proposed Ephemerizer, a centralized approach to secure data 
deletion. The goal of Ephemerizer is to find a balance between data availability and the 
ability to properly delete data. Users encrypt their data using a symmetric 
encryption scheme. Then they delegate key storage to a trusted third party. This 
third party destroys  cryptographic keys when they ``expire'', effectively making the original 
data unaccessible. Compared to~\cite{vanish}, \cite{CastellucciaCFK11}, as well as to our 
approach, Ephemerizer requires an always on-line, trusted third party.

\section{Conclusions}
\label{sec:conclusion}

In this paper, we have presented the first evaluation of covert ephemeral communication in NDN. 
Our techniques do not require \snr\ to exchange any packet directly. Rather, they rely on 
user-driven state on routers to publish and retrieve covert messages.
Messages published with our approach are ephemeral, i.e., they are automatically deleted from 
the network after a certain amount of time, without requiring any action from \sender\ or 
\receiver. Additionally, our delay-based techniques, messages {\em expire} immediately after 
being retrieved. 

Our techniques are based on fundamental components on NDN, and do not require ``abuse'' of 
application-layer protocols. In practice \snr\ only need access to non-popular content.

We performed experiments on a prototype implementation of our protocols. In particular, we 
measured the the bandwidth and robustness of our approaches on a local (LAN) setup and in a 
geographically distributed environment -- the official NDN testbed. Our experiments confirm 
that the techniques proposed in this paper provide high bandwidth and low error rate.

\balance
\bibliographystyle{abbrv}
{
\bibliography{references}
}

\end{document}